\documentclass[a4paper,11pt]{article}
\usepackage[dvipsnames]{xcolor}
\usepackage[utf8]{inputenc}
\usepackage[bbgreekl]{mathbbol}
\usepackage{geometry}

\usepackage{xcolor}
\usepackage{amsmath, array, amssymb, amsfonts,amsthm}
\usepackage{upgreek}
\usepackage{xfrac}
\usepackage[inline]{enumitem}
\setlist{nolistsep}
\usepackage[french, english]{babel}
\usepackage[all]{xy}
\usepackage{txfonts}  
\usepackage{sectsty}
\usepackage{booktabs}
\usepackage{caption}
\usepackage{dsfont}
\usepackage{mathtools}
\usepackage{slashed}
\usepackage[makeroom]{cancel}
\usepackage[hidelinks]{hyperref}
\usepackage{textcomp}
\usepackage{multicol}
\usepackage[title]{appendix}
\usepackage[square,numbers,sort,compress,semicolon,merge]{natbib}
\let\cite\citep 
\usepackage{hyperref}
\hypersetup{colorlinks=true, urlcolor=blue, citecolor=blue, linktoc=page}

\allowdisplaybreaks

\usepackage{tikz}
\usepackage{tikz-cd}
\usetikzlibrary{cd}
\tikzcdset{
arrow style=tikz,
diagrams={>={Straight Barb[scale=0.8]}}
}

\DeclareMathAlphabet{\mathpzc}{OT1}{pzc}{m}{it}

\usepackage{mathrsfs}
\usepackage{footmisc}
\usepackage{scrextend}

\makeatletter
\renewcommand*\env@matrix[1][\arraystretch]{%
  \edef\arraystretch{#1}%
  \hskip -\arraycolsep
  \let\@ifnextchar\new@ifnextchar
  \array{*\c@MaxMatrixCols c}}
\makeatother

\newcommand{\defeq}{\vcentcolon=}
\newcommand{\rdefeq}{=\vcentcolon}
\renewcommand\P{\mathcal{P}}

\newcommand\C{\mathcal{C}}

\newcommand\id{\textit{id}}

\renewcommand\H{\mathcal{H}}
\newcommand\A{\mathcal{A}}

\newcommand\vphi{\varphi}

\renewcommand\epsilon{\varepsilon}

\newcommand\rarrow{\rightarrow}

\newcommand\aut{\mathfrak{aut}}

\renewcommand\t{\tilde}

\renewcommand\b{\bar }
\newcommand\w{\wedge}

\newcommand\s{\sigma}
\newcommand\bs{\boldsymbol}

\renewcommand\-{^{-1}}
\newcommand\Ad{\text{Ad}}
\newcommand\ad{\text{ad}}
\renewcommand\id{\text{id}}

\makeatletter

\newcommand{\Rmnum}[1]{\expandafter\@slowromancap\romannumeral #1@}
\makeatother

\makeatletter
\newcommand{\leqnomode}{\tagsleft@true\let\veqno\@@leqno}
\newcommand{\reqnomode}{\tagsleft@false\let\veqno\@@eqno}
\makeatother

\DeclareMathOperator{\Diff}{Diff}
\newcommand\diff{\mathfrak{diff}}
\DeclareMathOperator{\Aut}{Aut}
\DeclareMathOperator{\Der}{Der}

\theoremstyle{definition}

\makeatother

\begin{document}


\title{ 
Note on the group of  vertical diffeomorphisms \\
of a principal bundle, \\
 \& its relation to the  Frölicher-Nijenhuis  bracket }
 \author{J. François}
\date{}

\maketitle
\begin{center}
\vskip -0.8cm
\noindent
Department of Mathematics \& Statistics, Masaryk University -- MUNI.\\
Kotlářská 267/2, Veveří, Brno, Czech Republic.\\[2mm]
 
Department of Philosophy -- University of Graz. \\
 Heinrichstraße 26/5, 8010 Graz, Austria.\\[2mm]
 
Department of Physics, Mons University -- UMONS.\\
 Service \emph{Physics of the Universe, Fields \& Gravitation}.\\
20 Place du Parc, 7000 Mons, Belgium.
 
%

\end{center}
%
\vspace{-3mm}

\begin{abstract}
The group of vertical diffeomorphisms of a principal bundle forms the generalised action Lie groupoid associated to the bundle. The former is generated by the group of maps with value in the structure group, which is also the group of bisections of the groupoid. 
The corresponding Lie algebra of general vertical vector fields is generated by maps  with value in the Lie algebra of the structure group. 
  The  bracket on these maps, induces by the bracket of vertical vector fields, is an  ``extended" bracket on gauge parameters: it has been introduced heuristically in physics, notably in the study asymptotic symmetries of gravity.
  Seeing the set of Lie algebra-valued maps  as sections of the action Lie algebroid associated to 
  the bundle,  the extended bracket is understood to be a Lie algebroid bracket on those sections.
  
Here, we  highlight that this bracket can also be seen to  arise from the Frölicher-Nijenhuis  bracket of  vector-valued differential forms.  
The benefit of this viewpoint is to insert this extended bracket within the general framework of derivations of forms on the bundle. Identities relating it to usual operations -- inner product, exterior and (Nijenhuis-) Lie derivative -- are immediately read as special cases of general results. 
We also look at the generalised gauge transformations induced by vertical diffeomorphisms, and discuss their peculiar features. In particular, locally, and contrary to standard gauge transformations arising from vertical bundle automorphisms, they are distinguishable from local gluings when iterated.  
Yet,  the gauge principle still holds.
\end{abstract}

\textbf{Keywords} : Bundle geometry, Frölicher-Nijenhuis bracket, Nijenhuis-Lie derivative, gauge field theory.

\vspace{-3mm}

\tableofcontents

\bigskip

 
\section{Introduction}  
\label{Introduction}  

In the some of the literature on gauge field theory and gravity, field-dependent gauge parameters or diffeomorphisms are often considered. These are needed, is it often argued, because of gauge-fixing or when boundary conditions are imposed that needs to be preserved by gauge symmetries. 
Consequently, a bracket extending the Lie algebra bracket of gauge parameters (the bracket of vector fields in the case of diffeomorphisms) is  introduced heuristically, so as to take into account their field-dependence. 
Such a bracket features often e.g. in the covariant phase space literature, notably in investigations of spacetime asymptotic symmetries (BMS and extensions). See e.g. \cite{Gomes-et-al2018, Freidel-et-al2021, Freidel-et-al2021bis, Speranza2022, Chandrasekaran-et-al2022} 
and references therein. Indeed this literature often traces back the introduction of this bracket to a paper by Barnich and Troessaert \cite{Barnich-Troessaert2009}. 
While these authors certainly deserve credit for independently coming up with it, the extended bracket was introduced earlier (maybe first) by Bergmann and Komar in   \cite{Bergmann-Komar1972}, and then further used by Salisbury and Sundermeyer in \cite{Salisbury-Sundermeyer1983}. 
%
To the best of our knowledge, the first clarification of its mathematical origin is to be found in \cite{Barnich2010}:  it is there understood as a Lie algebroid bracket. Indeed it can be shown that it corresponds essentially to the bracket of the  action Lie algebroid of vertical vector fields on the space of fields $\Phi$ of a gauge theory, seen as a principal bundle whose  structure group is the gauge group of the theory.

In this note, we highlight that it can also be understood as the simplest instance , in form-degree 0, 
of the the Frölicher-Nijenhuis (FN) bracket of vector-valued differential forms \cite{Kolar-Michor-Slovak} on $\Phi$. The field space Lie derivative along field-dependent vector fields, generating field-dependent gauge transformations, is then simply the Nijenhuis-Lie derivative. 

The observation holds in the simpler case of a finite dimensional principal bundle $P$ with structure group $H$. 
As~it is rarely articulated in basic treatments of the differential geometry of principal bundles, to advertise the fact to a broader audience, in section \ref{Groups of vertical and gauge transformations}-\ref{Linearisation and Lie algebraic structures} we  lay the case in some details from simple bundle theoretic considerations.
 We first describe the group $\Diff_v(P)$ of vertical diffeomorphisms of $P$, and characterise the corresponding group of generating $H$-valued maps on $P$. Then, we consider their respective Lie algebras.  We only briefly remind how these objects are naturally understood in Lie groupoid and Lie algebroid terms. 

Then, in section \ref{The Nijenhuis Lie derivative and the Frolicher Nijenhuis bracket on a principal bundle} we proceed to show that the bracket of general vertical vector fields is but an instance of the FN bracket, which indeed corresponds to the natural \emph{extended bracket} on the associated Lie algebra of Lie$H$-valued maps on $P$. 
 A benefit of viewing the extended bracket as a FN bracket, is that the many identities relating it to the exterior, Lie, and inner derivatives -- which are virtually always  derived heuristically in the physics literature -- are then understood as special cases of the general theory of derivations of the algebra $\Omega^\bullet(P)$, of which we remind the basics. This should at least streamlines technical aspects for the  physical applications (mainly to the covariant phase space literature). 
  
 In section \ref{General vertical transformations and the Nijenhuis-Lie derivative}, we consider the  issue of \emph{general vertical transformations} of forms on $P$, resulting from the action of $\Diff_v(P)$, and we show that  the infinitesimal version of those is given by the Nijenhuis-Lie derivative along general vertical vector fields. 
 A noticeable feature is that natural spaces of forms on $P$ -- tensorial forms and connections -- are not preserved by the action of $\Diff_v(P)$. 
In \ref{Local structure}, we consider the local picture, on the base manifold, and what it means for gauge theories living there: A peculiarity is that, contrary to local gauge transformations,  local generalised gauge transformations do distinguish themselves from local gluings of the bundle when iterated. 

As mentioned, these considerations apply with very little adjustments (mainly notational) to the infinite dimensional case of the field space $\Phi$ of gauge field theories, whose structure group can either be the internal gauge group $\H$ of the theory (i.e. of $P$) or $\Diff(M)$. We give a detailed account of the later case in a separate work \cite{Francois2023-a}, for which this paper can be considered a preparatory technical note.


\section{Vertical diffeomorphisms of a principal bundle and the Frölicher-Nijenhuis bracket}
\label{Vertical diffeomorphisms of a principal bundle and the Frolicher-Nijenhuis bracket}

\subsection{Groups of vertical and gauge transformations}
\label{Groups of vertical and gauge transformations}

Classical gauge gauge field theory is founded on the geometry of connections on fiber bundles, whose central objects are principal bundles. 
Consider a $H$-principal bundle $P$ over a base manifold $M$ (``spacetime" in gauge field theory), $H$ its structure (Lie) group.
 As a manifold, $P$ has a group of diffeomorphisms $\Diff(P)$, but as a bundle its maximal  group of transformations is the group of bundle automorphisms $\Aut(P)\defeq \left\{ \psi \in \Diff(P)\, |\,  \psi \circ R_h= R_h \circ \psi   \right\}$. The latter, commuting with the right action of $H$, preserve fibers, and thus project as diffeomorphisms of $M$. 
 
 The subgroup of \emph{vertical} diffeomorphisms of the $P$ is $\Diff_v(P) \defeq \left\{ \psi \in \Diff(P)\, |\, \pi \circ \psi = \pi \right\}$. These are diffeos $\psi$ moving along fibers, therefore  there is a unique smooth map 
$\gamma:P \rarrow H$ s.t. $\psi(p)=R_{\gamma(p)} p = p\gamma(p)$. 
In other words, to $\psi \in \Diff_v(P)$ corresponds a unique $\gamma \in C^\infty(P, H)$. 

This generalises  the subgroup of vertical automorphisms $\Aut_v(P)\defeq \left\{ \psi \in \Aut(P)\, |\,  \pi \circ \psi = \pi  \right\}$, isomorphic to the gauge group $\H\defeq \big\{ \gamma:P \rarrow H\, |\, R^*\gamma= h\- \gamma h \big\}$ via $\psi(p)=R_{\gamma(p)}\,p$: The constraint of $H$-equivariance of $\psi$  defining the group of bundle automorphisms $\Aut(P)$ is responsible for the specific equivariance required on elements $\gamma$ of the gauge group. 
Obviously then, $\Diff_v(P)  \cap \Aut(P) = \Aut_v(P)$. 

We may also notice that bundle automorphisms belong to the normaliser of vertical diffeomorphisms: Indeed, for $\vphi \in \Aut(P)$ and $\psi \in \Diff_v(P) \sim \gamma \in C^\infty(P, H)$ we have, 
\begin{align*}
\big( \vphi\- \circ \psi\circ\vphi \big) (p) = \vphi\- \circ \psi\big( \vphi(p) \big) = \vphi\-\big( R_{\gamma\left ( \vphi(p)  \right)} \, \vphi(p) \big) = R_{\gamma\left ( \vphi(p)  \right)}  \, \vphi\-\circ \vphi(p)
							= R_{\gamma\left ( \vphi(p)  \right)} \, p. 
\end{align*}
Therefore, $ \pi \circ \big( \vphi\- \circ \psi\circ\vphi \big) = \pi$, i.e. $\vphi\- \circ \psi\circ\vphi  \in \Diff_v(P)$. Since naturally a group is a subgroup of its normaliser, we have that, 
\begin{align}
\label{Normaliser}
N_{\Diff(P)}\big(\!\Diff_v(P) \big) = \Diff_v(P) \cup \Aut(P). 
\end{align}
As a special case, we have  $N_{\Diff(P)}\big(\!\Aut_v(P) \big) =  \Aut(P)$, i.e. the well-known fact that $ \Aut_v(P) \triangleleft  \Aut(P)$, which in turn gives us the short exact sequence (SES) of groups characteristic of a principal bundle:
\begin{align}
\label{SES-bundle}
\id  \rarrow \Aut_v(P)\simeq \H  \xrightarrow{\iota} \Aut(P)   \xrightarrow{\pi}  \Diff(M) \rarrow \id
\end{align}
The linearised version is the SES of Lie algebras, a.k.a. the Atiyah Lie algebroid associated to a principal bundle $P$.
\begin{align}
\label{SES-Atiyah-Alg}
0  \rarrow \aut_v(P)= \Gamma_H(VP)\simeq \text{Lie}\H  \xrightarrow{\iota} \aut(P)= \Gamma_H(TP)   \xrightarrow{\pi_*}  \diff(M)  \simeq \Gamma(TM)  \rarrow 0.
\end{align}
Here $ \Gamma_H(TP)\defeq \big\{ \bs X \in \Gamma(TP)\, |\, R_{h*}\bs X_{_p} =\bs X_{|ph} \big\}$ are the right-invariant vector fields of $P\,$: 
Indeed, for $\bs X\defeq \tfrac{d}{d\tau}\,  \psi_\tau \,\big|_{\tau=0}$ with flow $\psi_\tau \in \Aut(P)$, we have
\begin{align}
R_{h*} \bs X_{|p} \defeq \tfrac{d}{d\tau}\,  R_h \big( \psi_\tau(p)\big) \,\big|_{\tau=0}=\tfrac{d}{d\tau}\, \psi_\tau \big(  R_h\, p \big)\,\big|_{\tau=0} \rdefeq \bs X_{ph}. 
\end{align}
Naturally, $ \Gamma_H(VP)\defeq \big\{ \bs X \in \Gamma_H(TP)\, |\, \pi_*\bs X\equiv 0 \big\}$, while Lie$\H\defeq \big\{ X:P \rarrow \text{Lie}H\, |\, R^*X= \Ad_{h\-} X \big\}$. A $H$-invariant vertical vector field is  written as $X^v \defeq \tfrac{d}{d\tau}\,  \psi_\tau \,\big|_{\tau=0}$ with flow $\psi_\tau(p) =R_{\gamma_\tau(p)}\, p \in \Aut_v(P)$ and $X \defeq \tfrac{d}{d\tau}\, \gamma_\tau \,\big|_{\tau=0}$. In other words, elements of Lie$\H$ generate $H$-invariant fundamental vector fields. 
\medskip

The composition law in $\Diff_v(P)$ induces a peculiar composition law for the corresponding elements in $C^\infty(P, H)$. Indeed, for $\psi, \vphi \in \Diff_v(P)$ to which are respectively associated $\gamma, \eta\in C^\infty(P, H)$, we have $\vphi \circ \psi \in \Diff_v(P)$ (since $\pi \circ \vphi \circ \psi = \pi \circ \psi = \pi$) so one expects it has a corresponding generating element in  $C^\infty(P, H)$. 
The question is how the latter is expressed in terms of $\eta$ and $\gamma$. We have, 
\begin{equation}
\begin{aligned}
\label{comp-step}
(\varphi \circ \psi)(p) &= \varphi\left(  \psi(p) \right) = R_{\eta \left( \psi(p) \right)} \psi(p)= R_{\eta \left( R_{\gamma(p)} p \right)} R_{\gamma(p)} p,  \\
				&=R_{\gamma(p) \, \eta \left( R_{\gamma(p)} p \right)}\, p.
\end{aligned}
\end{equation}
So, 
\begin{align}
\label{comp}
\vphi \circ \psi \in \Diff_v(P) \quad \text{ is associated to } \quad \gamma\, ( \eta \circ R_\gamma)  \in C^\infty(P, H).
\end{align}
Observe that for vertical automorphisms $\psi, \vphi \in \Aut_v(P)$ to which  correspond  the gauge group elements $\gamma, \eta  \in \H$, we have $\eta \circ R_\gamma =R^*_\gamma \eta =\gamma\- \eta \gamma$. 
So~$\gamma ( \eta \circ R_\gamma) = \eta \gamma$, and \eqref{comp} reduces to the standard result $(\varphi \circ \psi)(p) =\big(R_{\eta\gamma}\big)p$. 
 
For the inverse diffeomorphism $\psi\-$,  $\psi\- \circ \psi =\id_P=\psi \circ \psi\-$, we must have  $\t\gamma \in C^\infty(P, H)$ s.t. $\psi\-(p)=R_{\t\gamma(p)} p$. According to \eqref{comp} we must then have that on the one hand,
\begin{align}
\label{i}
\psi\- \circ \psi =\id_P  \quad \text{ is associated to }  \quad \gamma ( \t\gamma \circ R_\gamma) =\id_H. 
\end{align}
And on the other hand, 
\begin{align}
\label{ii}
\psi \circ \psi\- =\id_P  \quad \text{ is associated to }  \quad \t\gamma ( \gamma \circ R_{\t\gamma}) =\id_H.
\end{align}
From \eqref{i} in particular, we get 
\begin{align}
\label{inverse}
\t\gamma \circ R_\gamma =\gamma\-_R.
\end{align}
 Observe that for $\gamma \in \H$, \eqref{ii} is $\gamma\t\gamma=\id_H$, i.e. $\t\gamma = \gamma\-_R$ for all $p\in P$. In particular for $ph$: $\t\gamma(ph)=\gamma_R(ph)\-=h\-\gamma_R(p)\- h = h\- \t\gamma(p) h$, i.e. $R^*_h \t\gamma = h\- \t\gamma h$. So, \eqref{i} is $\t\gamma\gamma=\id_H$ and $\t\gamma =\gamma\-_L$. Thus, $\t\gamma =\gamma\-$, as one would expect.

Following on \eqref{comp}, one shows that for $\psi_1, \psi_2, \psi_3 \in \Diff_v(P)$ associated respectively to $\gamma_1, \gamma_2, \gamma_3 \in  C^\infty(P, H)$:
\begin{align}
\label{comp-3}
\psi_3 \circ \psi_2 \circ \psi_1 \quad \text{ is associated to } \quad \gamma_1 ( \gamma_2 \circ R_{\gamma_1}) \big( \gamma_3 \circ R_{\gamma_1 (\gamma_2 \circ R_{\gamma_1} )} \big).
\end{align}
The pattern is clear, for $k$ iterated compositions we have, 
\begin{align}
\label{comp-k}
\circ_{i=1}^{k} \psi_i   \quad \text{ is associated to } \quad  \Pi_{i=1}^k t_i, \ \ \text{ with } t_i \defeq \gamma_i \circ R_{t_{i-1}}.
\end{align}
We therefore see that the composition law of $\Diff_v(P)$ doesn't give rise to a simple pointwise multiplicative group structure on the corresponding generating elements in $C^\infty(P, H)$, but rather to a recursive nested one involving the (yet unspecified) equivariance. 

This is but an example of the composition law of the infinite dimensional group of bisections of a Lie groupoid \cite{Mackenzie2005, Schmeding-Wockel2015} -- see also \cite{Maujouy2022} chap. 6 for a nice introduction. Indeed, the above can be reframed in the groupoid framework in the following way: 
One defines the generalised action groupoid $\Gamma  \rightrightarrows P$ with $\Gamma = P \rtimes \big(C^\infty(P, H) \simeq \Diff_v(P)\big)$, and with source and target maps 
$s: \Gamma \rarrow P$, $(p, \gamma\simeq \psi) \mapsto p$,  and 
$t: \Gamma \rarrow P$, $(p, \gamma\simeq \psi) \mapsto \psi(p)=R_{\gamma(p)}\, p $.
There is an associative composition law: $g\circ f \in \Gamma$ for $f,  g\in \Gamma$ defined whenever $t(f)=s(g)$.
 It naturally generalises the action groupoid $\b\Gamma =  P \rtimes H \rightrightarrows P$ associated with the right action of $H$ on $P$. The group of bisections $\mathcal B(\Gamma)$ of $\Gamma$ is the set of sections of  $s$ --  i.e. maps $\sigma : P \rarrow \Gamma$, $p\mapsto \s(p)=(p, \gamma \sim \psi)$  s.t. $s\circ \s =\id_P$ -- such that $t \circ \s: P\rarrow P$ is invertible: thus $t \circ \s \in \Diff(P)$. As a matter of fact, we here have indeed, for  $\s \in \mathcal B(\Gamma)$, that $t \circ \s=\psi=R_\gamma \in \Diff_v(P)\simeq C^\infty(P, H)$. 
The group law of bisections is defined as 
$\big(\s_2 \star \s_1\big)(p) \defeq \s_2\big( (t \circ \s_1 (p) \big) \circ \s_1(p)$. 
Composing this with the target map $t$ on the left reproduces precisely the peculiar group law \eqref{comp}. 
Thus, by a slight abuse of terminology, we may refer to $C^\infty(P, H)$ as the group of bisections of the generalised action groupoid $\Gamma$ associated to $P$. More simply, we may accurately just refer to it as the group of generating maps of $\Diff_v(P)$.


\subsection{Linearisation and Lie algebraic structures}
\label{Linearisation and Lie algebraic structures}

We now turn  to the question of linearisation, i.e. of describing the Lie algebra  $\diff_v(P)$: Consider a 1-parameter element $\psi_\tau \in \Diff_v(P)$ s.t. $\psi_\tau(p)=R_{\gamma_\tau(p)} p$ with $\gamma_\tau \in C^\infty(P, H)$. 
We have by definition,
\begin{align} 
\label{Gen-vert-element}
\tfrac{d}{d\tau} \gamma_\tau(p)\, \big|_{\tau=0} \rdefeq X(p) \in \text{Lie}H,
\end{align}
and $X\in C^\infty(P, \text{Lie}H)=\Omega^0(P, \text{Lie}H)$. 
By definition, an element of $\diff_v(P)$ is a \emph{general} vertical vector field:
\begin{align} 
\label{GenVectField}
\tfrac{d}{d\tau} \psi_\tau(p)\, \big|_{\tau=0}  &\rdefeq X(p)^v_{|p} \in V_pP, \\
								 &= pX(p), \notag
\end{align} 
 and $X^v \in \diff_v(P)  \simeq \Omega^0(P, VP)$. 
Now, 
\begin{align} 
\label{-X}
\tfrac{d}{d\tau}\,  \psi_\tau\- \circ \psi_\tau(p)\, \big|_{\tau=0}  &= \tfrac{d}{d\tau}\,  \psi_\tau(p)\, \tilde\gamma_\tau\big( \psi_\tau(p) \big)  \, \big|_{\tau=0}, \notag \\[1mm]
										0	\ \    &=  \tfrac{d}{d\tau}\,  \psi_\tau(p) \, \big|_{\tau=0}\,  \tilde\gamma_0\big( \psi_0(p) \big) + 
											         \psi_0(p)\,  \tfrac{d}{d\tau}\,  \tilde\gamma_\tau\big( \psi_\tau(p) \big)  \, \big|_{\tau=0},\notag \\[2mm]
					\Rightarrow \quad \tfrac{d}{d\tau}\,  \tilde\gamma_\tau\big( \psi_\tau(p) \big)  \, \big|_{\tau=0} &= -X(p)^v_{|p}. 		      
\end{align} 
The bracket of vertical vector fields \eqref{GenVectField} must be a vertical vector field itself. 
We want the expression of its generating element in $\C^\infty(P, \text{Lie}H)$.
Consider $X^v$ generated by $\psi_\tau \in \Diff_v(P)$ with $X(p)=\tfrac{d}{d\tau} \gamma_\tau(p)\, \big|_{\tau=0} \in C^\infty(P, \text{Lie}H)$, 
and  $Y^v$ generated by $\vphi_\tau \in \Diff_v(P)$ with $Y(p)=\tfrac{d}{d\tau} \eta_\tau(p)\, \big|_{\tau=0} \in C^\infty(P, \text{Lie}H)$. Their bracket is: 
\begin{align*}
\big[ X(p)^v, Y(p)^v\big]_{|p} \defeq&\,  \tfrac{d}{d\tau}\tfrac{d}{ds}\,  \psi_\tau\- \circ \vphi_s \circ \psi_\tau(p)\, \big|_{s=0} \big|_{\tau=0}, \\
					     =&\,  \tfrac{d}{d\tau}\tfrac{d}{ds}\, \psi_\tau(p) \cdot \eta_s\big( \psi_\tau(p) \big) \cdot \tilde\gamma_\tau\left( \vphi_s\big( \psi_\tau(p) \big) \right)  \big|_{s=0} \big|_{\tau=0}, \quad \text{by \eqref{comp-3},}\\
					     =&\, \tfrac{d}{d\tau} \bigg\{  \psi_\tau(p)\, \underbrace{\tfrac{d}{ds}\, \eta_s\big( \psi_\tau(p) \big)\, \big|_{s=0}}_{Y\big( \psi_\tau(p) \big)} \cdot \underbrace{\tilde\gamma_\tau\bigg( \underbrace{\vphi_0\big( \psi_\tau(p) \big)}_{\psi_\tau(p)} \bigg) }_{\gamma_\tau(p)\- \text{ by \eqref{inverse}}}
					                                         \ + \      \psi_\tau(p)\cdot \underbrace{\eta_0  \big( \psi_\tau(p) \big)}_{\id_H} \,   \underbrace{\tfrac{d}{ds}\,   \tilde\gamma_\tau\left( \vphi_s\big( \psi_\tau(p) \big) \right)   \big|_{s=0} }_{\big[ Y^v(\tilde\gamma_\tau) \big]\big(\psi_\tau(p)\big)}    \bigg\}  \big|_{\tau=0}, \\
					      =&\,  \tfrac{d}{d\tau} \bigg\{  p\, \Ad\big(\gamma_\tau(p)\big) Y\big( \psi_\tau(p) \big) + \psi_\tau(p)\cdot \big[ Y^v(\tilde\gamma_\tau) \big]\big(\psi_\tau(p)\big)  \bigg\}  \big|_{\tau=0}, \\
					        =&\, p  \bigg\{  \tfrac{d}{d\tau}\, \Ad\big(\gamma_\tau(p)\big) Y(p)\,  \big|_{\tau=0} + \Ad\big(\underbrace{\gamma_0(p)}_{\id_H}\big) \tfrac{d}{d\tau}\,  Y\big( \psi_\tau(p) \big)\, \big|_{\tau=0} \, + \, \tfrac{d}{d\tau}\,  \big[ Y^v(\tilde\gamma_\tau) \big]\big(\psi_\tau(p)\big)  \,\big|_{\tau=0}	 \bigg\}, \\
					         =&\, p  \bigg\{ \ad\big( X(p) \big) Y(p) + \big[X^v(Y) \big](p) + \big[Y^v(-X) \big](p)  \bigg\}, \\
					          =&\!: [X(p), Y(p)]^v_{|p} + \big\{ [X^v(Y)](p) \big\}^v_{|p} - \big\{ [Y^v(X)](p) \big\}^v_{|p}.
\end{align*}  
Thus we obtain, for $X^v, Y^v \in  \diff_v(P)$ 
generated by $X, Y \in C^\infty(P, \text{Lie}H)$:
\begin{equation}
  \begin{aligned}
\big[ X^v, Y^v\big] 
			    = &\, \{X, Y\}^v,\\[1mm]
\text{with} \quad \{X, Y\}  \defeq&\,   [X, Y]_{\text{\tiny{Lie$H$}}} +  X^v(Y) - Y^v(X).      \label{ext-bracket}
\end{aligned}
\end{equation}
The sought after generating element of the bracket of general vertical vector fields is then $\{X, Y\} \in C^\infty(P, \text{Lie}H)$. 
This~\emph{extended} bracket $\{\ ,\, \}$, reflecting the recursive nested structure \eqref{comp-3}-\eqref{comp-k} on the elements of $C^\infty(P, H)$ generating $\Diff_v(P)$,  is manifestly antisymmetric in $X, Y$ and satisfies the Jacobi identity (as is easily proven). 
Therefore,  $C^\infty(P, \text{Lie}H)$ equiped with $\{\ ,\, \}$ is a Lie algebra: We have thus the Lie algebra isomorphism $\diff_v(P) \simeq C^\infty(P, \text{Lie}H)$ (respective brackets tacitly understood). 
It follows in particular that the Lie derivative and inner product along $\diff_v(P)$ satisfy:
\begin{equation}
  \begin{aligned}
   \label{Id-ext-bracket}
 [L_{{ X}^v},  L_{ { Y}^v }]  &=  L_{[ { X}^v,  { Y}^v]}=  L_{\{ { X},  { Y}\}^v }, \\
 [L_{{ X}^v},  \iota_{ { Y}^v }] &=  \iota_{[ { X}^v,  { Y}^v]}=  \iota_{\{ { X},  { Y}\}^v }.
 \end{aligned}
\end{equation}

Notice that for $X,Y \in$ Lie$\H$, we have $R^*_h Y=\Ad_{h\-} Y$, so $X^v(Y)=[Y, X]_{\text{\tiny{Lie$H$}}}$. Therefore, the extended bracket reduces to $\{X, Y\} = -[X, Y]_{\text{\tiny{Lie$H$}}}$ and $[X^v, Y^v]=(-[X, Y]_{\text{\tiny{Lie$H$}}})^v$ as it should. This is the standard result that the ``verticality map" $|^v : \text{Lie}\H \rarrow \aut_v(P)\simeq\Gamma_H(VP)$, $X \mapsto X^v$ is an anti-isomorphism. 

We may also observe that given a connection $\omega$ on $P$, satisfying $\omega(X^v)=X$, its curvature is a tensorial 2-form that can be expressed via Cartan's structure equation $\Omega=d\omega +\tfrac{1}{2}[\omega, \omega]_{\text{\tiny{Lie$H$}}} $. And indeed one may check (in the very  process of proving that equation) that, on $X^v, Y^v \in \diff_v(P)$:
\begin{align}
\Omega(X^v, Y^v) &= d\omega (X^v, Y^v) + [\omega (X^v), \omega(Y^v)]_{\text{\tiny{Lie$H$}}}, \notag \\
			    &= X^v \cdot \omega(Y^v) - Y^v\cdot \omega(X^v) - \omega([X^v, Y^v]) \ +\ [X, Y]_{\text{\tiny{Lie$H$}}},\notag \\
			    &= X^v (Y) - Y^v(X) - \omega(\{X, Y\}^v) \ +\ [X, Y]_{\text{\tiny{Lie$H$}}} \equiv 0.
\end{align}

In echo to the end of the previous section, where we reminded the Lie groupoid structure behind $\Diff_v(P)$, let us briefly recall the Lie algebroid picture of the above \cite{Mackenzie2005, Crainic-Fernandez2011}. Associated to the action of Lie$H$ on $P$, i.e. to the Lie algebra morphism $\alpha = |^v: \text{Lie}H \rarrow \Gamma(VP) \subset \Gamma(TP)$, is the action (or transformation) Lie algebroid $A=P \rtimes \text{Lie}H \rarrow P$ with anchor $\rho : A \rarrow VP \subset TP$, $(p, X) \mapsto \alpha(X)_{|p}=X^v_{|p}$, inducing the Lie algebra morphism $\t \rho: \Gamma(A) \rarrow \Gamma(VP) \subset \Gamma(TP)$. 
Here, the space of sections $\Gamma(A)=\{ P \rarrow A, p \mapsto \big(p, X(p)\big) \}$ is naturally identified with the space $C^\infty(P, \text{Lie}H)$,\footnote{Sections of $\Gamma(A)$ are the graphs of elements of $C^\infty(P, \text{Lie}H)$.} so~that $\t \rho=\alpha=|^v$.  
The Lie algebroid bracket $[\ \,,\  ]_{\text{\tiny{$\Gamma(A)$}}}$ is 
uniquely determined by the Leibniz condition $[X, f Y]_{\text{\tiny{$\Gamma(A)$}}}=f[X, Y]_{\text{\tiny{$\Gamma(A)$}}} + \t \rho(X) f \cdot Y$, for $f\in C^\infty(P)$,
and the requirement that  $[\ \,,\ ]_{\text{\tiny{$\Gamma(A)$}}} = [\ \, , \  ]_{\text{\tiny{Lie$H$}}}$ on constant sections.
It is found to be: $[X, Y]_{\text{\tiny{$\Gamma(A)$}}} = [X, Y  ]_{\text{\tiny{Lie$H$}}} + \t \rho (X) Y - \t\rho(Y) X$.
It indeed reproduces \eqref{ext-bracket} above, obtained as the result of a computation which amounts to showing explicitly  
that $\t\rho=|^v$ is indeed a Lie algebra morphism. 

Quite intuitively, the action Lie algebroid $A$ is the limit of the action Lie groupoid $\Gamma$, when the target and source maps are infinitesimally close. Hence $C^\infty(P, \text{Lie}H)\simeq \Gamma(A)$ is the Lie algebra of the group (of bisections) $C^\infty(P, H)$, as we've shown  explicitly above. 

In the next section, we show that the extended bracket \eqref{ext-bracket}  
can also be understood as the degree 0 of the the Frölicher-Nijenhuis bracket of vector-valued forms on $P$, while the Lie derivative along $\diff_v(P)$    \eqref{Id-ext-bracket} is but a case of the Nijenhuis-Lie derivative. 
As we are about to see, the benefit of this viewpoint is that it nests this bracket within a general theory of derivations of the algebra of differential forms $\Omega^\bullet(P)$: As such, its links to the usual operations, such as the exterior derivative and inner product, are readily understood as simple instances of general results. 

In the physics literature (e.g.  on covariant phase space methods), some computational efforts are often spent on re-deriving various identities relating the extended bracket to the familiar Cartan calculus of forms.
On account of  the viewpoint highlighted here, these efforts can be spared. One needs only to point to the established mathematical literature.


\subsection{The Nijenhuis-Lie derivative and the Frölicher-Nijenhuis bracket on a principal bundle}
\label{The Nijenhuis Lie derivative and the Frolicher Nijenhuis bracket on a principal bundle}

  We refer to \cite{Kolar-Michor-Slovak} (chap. II, section 8) for a systematic presentation of the following notions (on a generic smooth manifold) and for proofs of the relations displayed.

As a manifold, $P$ has a space of  differential forms $\Omega^\bullet(P)$.  Its space of derivations forms a graded Lie algebra
$\Der_\bullet \big(\Omega^\bullet(P) \big)=\bigoplus_k \Der_k \big(\Omega^\bullet(P) \big)$,  with graded bracket  $[D_k, D_l]=D_k \circ D_l - (-)^{kl} D_l \circ D_k$, with $D_i \in \Der_i\big(\Omega^\bullet(P) \big)$.  

 The de Rham complex of $P$ is $\big( \Omega^\bullet(P);  d  \big)$ with $ d \in \Der_1$ the de Rham (exterior) derivative, which is nilpotent -- $ d ^2 =0=\sfrac{1}{2}[ d,  d]$ -- and defined via the Koszul formula.  Given the exterior product $\w$ defined as usual on scalar-valued forms, we have that  
 $\big( \Omega^\bullet(P, \mathbb K), \w,  d \big)$ is a differential graded algebra.\footnote{The exterior product can also be defined on the space $\Omega^\bullet(P, \sf A)$ of variational differential forms with values in an algebra $(\sf A, \cdot)$, using the product in $\sf A$ instead of the product in the field $\mathbb K$. So  $\big( \Omega^\bullet(P, {\sf A}), \w,  d \big)$ is a again a differential graded algebra. On the other hand, an exterior product cannot be defined on $\Omega^\bullet(P, \bs V)$ where $\bs V$ is merely a vector space.}

One may define the subset of vector-field valued differential forms $\Omega^\bullet(P, TP)=\Omega^\bullet(P) \otimes TP$. Then, the subalgebra of 
 ``\emph{algebraic}" derivations is defined as $D_{|\Omega^0(P)}=0$, they have the form $\iota_{ K} \in \Der_{k-1}$ for $ K \in \Omega^{k}(P, TP)$, with $\iota$ the inner product. 
 For $\omega  \otimes   X \in \Omega^\bullet(P, TP)$ we have : $\iota_{K}( \omega  \otimes   X)  := \iota_{ K} \omega\otimes   X = \omega \circ  K \otimes   X$.
 On $\Omega^\bullet(P, TP)$,  the \emph{Nijenhuis-Richardson bracket} (or \emph{algebraic} bracket) is defined by: 
\begin{align}
\label{NR-bracket} 
 [ K,  L]_{\text{\tiny{NR}}}:=\iota_{ K}  L -(-)^{(k-1)(l-1)} \, \iota_{ L}  K. 
 \end{align}
 It generalises the inner contraction of a form on a vector field, and  it makes the map: 
  \begin{align}
  \iota: \Omega^\bullet(P, TP) &\rarrow \Der_\bullet \big(\Omega^\bullet(P) \big) \\ 
  					 K &\mapsto  \iota_{ K}         \notag 
  \end{align}
a graded Lie algebra morphism, since:
\begin{align}
\label{NR-bracket-id}
[\iota_{ K}, \iota_{ L}]=\iota_{[ K,  L]_{\text{\tiny{NR}}}}. 
\end{align}
The \emph{Nijenhuis-Lie derivative} is the map,  
\begin{equation}  
\label{NL-derivative}
  \begin{aligned}    
   L:=[\iota,  d] : \Omega^\bullet(P, TP) &\rarrow \Der_\bullet \big(\Omega^\bullet(P) \big) \notag\\
  					 K &\mapsto   L_{ K}:= \iota_{ K}  d - (-)^{k-1}  d  \iota_{ K}      
  \end{aligned}
  \end{equation}     
We have $ L_{ K} \in \Der_k$ for $ K \in \Omega^{k}(P, TP)$.  It generalises the Lie derivative along vector fields, $ L_{X} \in \Der_0$. 
It is such that $[ L_{ K},  d]=0$, and it is a  morphism of graded Lie algebras: 
\begin{align}
\label{NL-deriv-morph}
[ L_{ K},  L_{ J}]= L_{[ K,  J]_{\text{\tiny{FN}}}},
\end{align}
 where $[ K,  J]_{\text{\tiny{FN}}}$ is the \emph{Frölicher-Nijenhuis bracket}.  Explicitely, for $ K = \mathsf K \otimes  X \in \Omega^k(P, TP)$ and   
 $ J = \mathsf J \otimes  Y \in \Omega^l(P, TP)$, it is: 
  \begin{align}
  \label{FN-bracket}
  [ K,  J]_{\text{\tiny{FN}}} \defeq \mathsf K \w \mathsf J \otimes [ X,  Y]  + \mathsf K \w  L_{X} \mathsf J \otimes  Y
  														  -   L_{Y} \mathsf K \w J \otimes  X
														  + (-)^k \big(  d\mathsf K \w \iota_{X} \mathsf J \otimes  Y
														                       + \iota_{Y} \mathsf K \w d\mathsf J \otimes  X   \big).
  \end{align}
We further have the relations: 
\begin{equation}
  \begin{aligned}
  \label{Relations-L-iota}
[ L_{ K}, \iota_{ J}] &= \iota^{}_{ [ K,  J]_{\text{\tiny{FN}}} } - (-)^{k(l-1)}  L_{(\iota_{ K}   J)},  \\
[\iota_{ J}, \bs L_{ K}] &=  L_{(\iota_{ K}  J)} +(-)^k \, \iota^{}_{ [ J,  K]_{\text{\tiny{FN}}} }. 
  \end{aligned}
  \end{equation}

  \medskip
  
The FN bracket \eqref{FN-bracket} reproduces as a special case the extended bracket \eqref{ext-bracket}.
Indeed, specialising first \eqref{NR-bracket}  in degree $0$, for  $ f = \mathsf f \otimes  X$ and $ g = \mathsf g \otimes  Y \in \Omega^0(P, TP)$, we have: 
\begin{equation}
  \begin{aligned}
  \label{NR-bracket-special}
  [ f, {dg}]_{\text{\tiny{NR}}} &= \iota^{}_{ f } {dg} -(-)^0  \cancel{\iota^{}_{{dg}}  f} = \big(  \mathsf f \w \iota_{X}  d \mathsf g\big)\otimes  Y 
  = \big(\mathsf  f \w  L_{X} \mathsf  g\big)\otimes  Y,   \\
   [{df}, {g}]_{\text{\tiny{NR}}} &= \cancel{  \iota^{}_{{df}} {g} }  -(-)^0  \iota^{}_{{g}} {d f} = -  [{g}, {df}]_{\text{\tiny{NR}}} =-  \big( \mathsf g \w  L_{Y} \mathsf f\big)\otimes  X.
  \end{aligned}
  \end{equation}
 So that   \eqref{FN-bracket} is:
    \begin{align}
    \label{FN-bracket-special}
     [ f,  g]_{\text{\tiny{FN}}} &=\mathsf f \w\mathsf g \otimes [ X,  Y]  + f \w  L_{X}\mathsf g \otimes  Y
  														  -   L_{Y} f \w \mathsf g \otimes  X, \notag \\
						&= \mathsf f \w\mathsf  g \otimes [ X,  Y] +  [ f, {dg}]_{\text{\tiny{NR}}}  -  [{g}, {df}]_{\text{\tiny{NR}}},
    \end{align}
 and \eqref{NL-deriv-morph}-\eqref{Relations-L-iota} reduces to:
\begin{equation}
\begin{aligned}
\label{FN-NL-der}
   [ L_{ f},  L_{ g}]&= L_{[ f,  g]_{\text{\tiny{FN}}}},\\
   [ L_{ f}, \iota_{ g}] &= \iota^{}_{ [ f,  g]_{\text{\tiny{FN}}} }. 
\end{aligned}      
\end{equation}
Now, the map  $|^v : C^\infty\big(P, \text{Lie}H\big) \rarrow  \Gamma(VP)$,  ${ X} \mapsto {  X}^v$, allows to think of ${ X}^v \in \diff_v(P)$ as a (vertical) vector-valued 0-form on $P$, i.e. $ { X}^v \in \Omega^0(P, VP) \subset \Omega^\bullet(P, TP)$. 
Explicitly, one may write $X^v=X^a \otimes (\tau_a)^v$, for $\{\tau_a\}$ a basis of Lie$H$, and $(\tau_a)^v$ the induced basis of fundamental vector fields for $VP \subset TP$.
Therefore, the Nihenhuis-Richardson and Frölicher-Nijenhuis brackets naturally apply:  specialising further 
eq \eqref{FN-bracket-special}, 
we have on the one hand
\begin{equation}
\label{NR-bracket-field-dep-diff}
\begin{split}
 [{ X}^v,  d { Y}^v ]_{\text{\tiny{NR}}} &= X^a \wedge \iota_{(\tau_a)^v}dY^b \otimes (\tau_b)^v  = X^a  \, L_{(\tau_a)^v}Y^b\otimes (\tau_b)^v = L_{X^v}Y^b \otimes (\tau_b)^v \\
                                                          &\rdefeq (L_{X^v}Y)^v = \{ { X}^v({ Y})\}^v, 
\end{split}                                                     
\end{equation}
and similarly, $-[{ Y}^v,  d { X}^v ]_{\text{\tiny{NR}}} = -\{ \iota_{{ Y}^v} {d}{ X} \}^v = - \{ { Y}^v({ X})\}^v$. 
On the other hand
 \begin{equation}
 X^a \wedge Y^b \otimes [(\tau_a)^v ,(\tau_b)^v]_{\text{{\tiny $\Gamma(TP)$}}}= X^a \, Y^b \otimes ([\tau_a, \tau_b]_{\text{{\tiny Lie$H$}}})^v= ([X, Y]_{\text{{\tiny Lie$H$}}})^v,
\end{equation}
where we use the fact that $|^v: \text{Lie}H \rarrow \Gamma(VP)$ is a Lie algebra morphism.
 So,  the FN bracket  \eqref{FN-bracket-special} for ${ X}^v, {  Y}^v \in \Omega^0(P, VP)$  is:
 \begin{align}
 \label{FN-bracket-field-dep-diff=BT-bracket}
   [{ X}^v,  { Y}^v ]_{\text{\tiny{FN}}} &=\big( [{ X}, { Y}]_{\text{{\tiny Lie$H$}}}\big)^v + [{ X}^v,  d{ Y}^v]_{\text{\tiny{NR}}}  - [{ Y}^v,  d\bs{ X}^v]_{\text{\tiny{NR}}}, \notag \\
    &=  \big(  [{ X}, { Y}]_{\text{{\tiny Lie$H$}}} +  { X}^v({ Y}) -   { Y}^v({ X}) \,\big)^v   = \{ { X},  { Y}\}^v. 
 \end{align}
 Which  reproduces the extend bracket \eqref{ext-bracket} on $C^\infty(P, \text{Lie}H)$.
 Then of course we have the following special cases of  \eqref{NR-bracket-id}, \eqref{NL-deriv-morph}-\eqref{Relations-L-iota}/\eqref{FN-NL-der}, among derivations in $\Der^\bullet \big(\Omega^\bullet(P) \big)$: 
\begin{equation}
\begin{aligned}
 [\iota_{{ X}^v}, \iota_{\{{d Y}\}^v}] &= \iota^{}_{ [{ X}^v,  d { Y}^v ]_{\text{\tiny{NR}}} } = \iota^{}_{ \{ \iota_{{ X}^v} {d}{ Y} \}^v },  \\
  [ L_{\bs{ X}^v}, \iota_{ { Y}^v }] &= \iota^{}_{[{ X}^v,  { Y}^v ]_{\text{\tiny{FN}}}}=  \iota_{\{ { X},  { Y}\}^v },  \\
 [ L_{\bs{ X}^v},  L_{ { Y}^v }] &=  L_{[{ X}^v,  { Y}^v ]_{\text{\tiny{FN}}}} 
 										    =  L_{\{ { X},  { Y}\}^v }.     \label{NL-FN-der-ext-bracket}
\end{aligned}      
\end{equation}
 As one should expect, \eqref{NL-FN-der-ext-bracket} reproduces \eqref{Id-ext-bracket}. 
 \medskip

 All of the above applies, mutatis mutandis, to the the infinite dimensional field space $P=\Phi$ of a gauge field theory, which can be seen (under some adequate mild restrictions) as a principal bundle with structure group $H=\H$, the internal gauge group a theory, or  $H=\Diff(M)$. 
 
To the best of our knowledge, the bracket \eqref{ext-bracket}-\eqref{FN-bracket-field-dep-diff=BT-bracket}  has been first introduced in physics by 
 Bergmann \& Komar \cite{Bergmann-Komar1972}, for the field space $\Phi=\{g_{\mu\nu}\}$, and Salisbury \& Sundermeyer \cite{Salisbury-Sundermeyer1983} in their investigations of the largest possible  covariance group of General Relativity -- see eq.(3.1)-(3.2) in \cite{Bergmann-Komar1972} and eq.(2.1) in  \cite{Salisbury-Sundermeyer1983}. It was latter reintroduced by Barnich \& Troessaert in their study  \cite{Barnich-Troessaert2009} of asymptotic symmetries of gravity in the flat limit at null infinity. This Bergmann-Komar-Salisbury-Sundermeyer (BKSS) bracket 
 for \emph{field-dependent vector fields} $\xi, \zeta  \in C^\infty\big( \Phi, \diff(M) \big)$, 
 $ \{ \xi,   \zeta\}_{\text{\tiny{BKSS}}} =- \{ \xi, \zeta  \}$, 
 appears in eq.(8) in \cite{Barnich-Troessaert2009} and more recently e.g. in eq.(3.3) in \cite{Freidel-et-al2021}, eq.(2.12) in \cite{Freidel-et-al2021bis}, eq.(2.21) in \cite{Gomes-et-al2018}, eq.(2.3) in \cite{Chandrasekaran-et-al2022}, eq.(1.1) in \cite{Speranza2022}.  
 As this references show, this bracket is commonly used in the covariant phase space literature concerned with the covariant symplectic structure of gravity and with the analysis of its asymptotic symmetries (BMS and extensions).

The extended bracket for field-dependent vector fields (or gauge parameters) can well be  understood in terms of action Lie algebroid bracket on $\Gamma(\A)\simeq C^\infty \big(\Phi, \diff(M)\big)$ for $\A= \Phi \rtimes  \diff(M) \rarrow \Phi$ (see \cite{Barnich2010} eq. (28)), itself arising from the group of bisections $C^\infty \big(\Phi, \Diff(M)\big)$ of the action groupoid $\Gamma = \Phi \rtimes \big( \bs\Diff_v(\Phi) \simeq C^\infty(\Phi, \Diff(M)\big )$.
   But viewing it as the degree 0 FN bracket on $\Phi$ allows to take advantage of the general theory of derivations of $\Omega^\bullet(\Phi)$, so has to obtain e.g. identities such as \eqref{NL-FN-der-ext-bracket} effortlessly.
 This might be useful for the applications pursued in the mentioned literature, where regularly some efforts are invested in re-deriving such identities (and others similar). 
 For example,  \eqref{NL-FN-der-ext-bracket} reproduces e.g. eq.(3.1) in \cite{Freidel-et-al2021}, or eq.(2.13) in \cite{Freidel-et-al2021bis}. To their credit, the authors of \cite{Chandrasekaran-et-al2022} essentially exploit  this viewpoint,  pointing to the  reference \cite{Kolar-Michor-Slovak}. Unfortunately, if they mention the Nijenhuis-Richardson bracket,  they  stop short from naming explicitly 
 the FN bracket as a key relevant mathematical notion.

We may then summarise the different ways in which the extended bracket can be understood, via the following Lie algebra isomorphism and equalities:
\begin{equation}
\Big( C^\infty(P, \text{Lie}H); \{\, \ , \ \}\Big) = \Big( \Gamma(A); [\, \ , \ ]_{\text{\tiny{$\Gamma(A)$}}} \Big) \simeq \Big( \diff_v(P); [\, \ , \ ]_{\text{\tiny{$\Gamma(TP)$}}} \Big) = \Big( \Omega^0(P, VP) ; [\, \ , \ ]_{\text{\tiny{FN}}} \Big) 
\end{equation}

\section{General vertical transformations and the Nijenhuis-Lie derivative}
\label{General vertical transformations and the Nijenhuis-Lie derivative}

Standard gauge transformations of a form $\beta \in \Omega^\bullet(P)$ are defined as \emph{vertical transformations given by the action of $\Aut_v(P)$ via pullback}, expressible in terms of the associated elements in the gauge group $\H$ of $P$\,: $\beta^\gamma\!\defeq \psi^*\beta$. Geometrically, this is computed by relying on the duality pullback/pushforward, ${\beta^\gamma}_{|p}(\bs X_{|p})\defeq \psi^*\beta_{|\psi(p)}(\bs X_{|p})= \beta_{|\psi(p)}(\psi_*\bs X_{|p})$, and one then only needs to find the result of the pushforward $\psi_*\bs X$ of any vector field $\bs X \in \Gamma(TP)$ by a $\psi \in \Aut_v(P)$. This is a standard computation, which turns out not to depend on the $H$-equivariance of $\psi$, so the result holds as well for $\psi \in \Diff_v(P)$. This will thus allow to define
\emph{general vertical transformations} of $\beta$ as the action of $\Diff_v(P)$ by pullback,\footnote{In \cite{Nishimura2018}, the group of bisections $\mathcal B(\Gamma)\simeq \Diff_v(P)\simeq C^\infty(P, H)$ of the action Lie groupoid $\Gamma$  is called ``\emph{generalised gauge transformations}". } expressible in terms of the associated elements of $C^\infty(P, H)$. 
The linearisation of these is treated next, and seen to be exactly given by the Nijenhuis-Lie derivative.

\subsection{Finite vertical and gauge transformations}
\label{Finite vertical and gauge transformations}

As a useful lemma, let us first derive the pushforward by the right $H$-action of  $X^v \in \diff_v(P)$ generated by $X= \tfrac{d}{d\tau}\, \gamma_\tau \, \big|_{\tau=0} \in C^\infty(P, \text{Lie}H)$, for $\gamma_\tau \in C^\infty(P, H)$:
\begin{equation}
\label{Gen.VVF.Equiv}
\begin{aligned}
R_{h*} X(p)^v_{|p}&= R_{h*}  \tfrac{d}{d\tau}\, R_{\gamma_\tau(p)} \, p \, \big|_{\tau=0} =  \tfrac{d}{d\tau}\, R_{h}  R_{\gamma_\tau(p)} \, p \, \big|_{\tau=0} 
			      = \tfrac{d}{d\tau}\,  R_{\gamma_\tau(p)h} \, p \, \big|_{\tau=0} =  \tfrac{d}{d\tau}\,  R_{ h\- \gamma_\tau(p)h} \,    R_{h}\, p \, \big|_{\tau=0},  \\
		            &\rdefeq \big(\Ad_{h\-}X(p)  \big)^v_{|ph} 
\end{aligned}
\end{equation}
Remark that, for $X \in$ Lie$\H$, we have $R^*_h X =\Ad_{h\-} X $, so $R_{h*} X(p)^v_{|p}=  X(ph) ^v_{|ph}$. That is, $X^v$ is a right-invariant vertical vector field, those forming (as already observed) the Lie algebra of vertical automorphisms \mbox{$\Gamma_H(VP)\simeq \aut_v(P)$}.

Now, the classic computation:
 For $\bs X \in \Gamma(TP)$ with flow $\phi_\tau$, and  given $\psi \in \Diff_v(P)$ to which corresponds $\gamma \in C^\infty(P, H)$, we have,
\begin{align*}
\psi_*\bs X_{|p} \defeq \tfrac{d}{d\tau}\, \psi\big( \phi_\tau(p) \big) \, \big|_{\tau=0} 
                          &=  \tfrac{d}{d\tau}\,  R_{\gamma\big(  \phi_\tau(p) \big)}\, \phi_\tau(p)   \, \big|_{\tau=0}, \\
                          &= \tfrac{d}{d\tau}\,   R_{\gamma\big(  \phi_\tau(p) \big)}\,  p   \, \big|_{\tau=0}   +  \tfrac{d}{d\tau}\,  R_{\gamma(p)}\, \phi_\tau(p)   \, \big|_{\tau=0}, \\
                          &= \tfrac{d}{d\tau}\,   R_{ \gamma(p)\gamma(p)\- \gamma\big(  \phi_\tau(p) \big)}\,  p   \, \big|_{\tau=0}   +  R_{\gamma(p)*} \bs X_{|p} ,\\
                          &= \tfrac{d}{d\tau}\,   R_{\gamma(p)\- \gamma\big(  \phi_\tau(p) \big)}\,  \underbrace{R_{\gamma(p)}\, p}_{\psi(p)}   \, \big|_{\tau=0}   +  R_{\gamma(p)*} \bs X_{|p}.
\end{align*}
The first term manifestly is a vertical vector field at $\psi(p)$, one needs only to find a nice way to write the associated  Lie algebra element (corresponding to the curve $\gamma(p)\- \gamma\big(  \phi_\tau(p) \big)$ through $e=\id_H$ in $H$). Let us first notice that, since $\gamma: P \rarrow H$, generically $\tfrac{d}{d\tau}\, \gamma\big(  \phi_\tau(p) \big)\, \big|_{\tau=0} = d\gamma_{|p}\big( \bs X_{|p}\big) = \gamma_*\big( \bs X_{|p}\big) \in T_{\gamma(p)}H$. While the Maurer-Cartan form on $H$ is, 
\begin{align*}
\theta_{\text{\tiny{MC}},\,  |h}\defeq L_{h\-*} : T_hH &\rarrow T_eH \rdefeq \text{Lie}H, \\
						                 \upchi_{|h} &\mapsto L_{h\-*}\, \upchi_{|h}. \\[1mm]
   \text{So,}  \ \quad  \big[ d\gamma_{|p} \big( \bs X_{|p}\big) \big]_{|\gamma(p)}   &\mapsto L_{\gamma(p)\-*}  \big[ d\gamma_{|p} \big( \bs X_{|p}\big) \big]_{|\gamma(p)} 
   																         =  \big[  \gamma(p)\- d\gamma_{|p} \big( \bs X_{|p}\big) \big]_{|e}, \\
													& \hspace	{.6cm}  L_{\gamma(p)\-*}   \tfrac{d}{d\tau}\, \gamma\big(  \phi_\tau(p) \big)\, \big|_{\tau=0}      = \tfrac{d}{d\tau}\, \gamma(p)\-\gamma\big(  \phi_\tau(p) \big)\, \big|_{\tau=0}.
\end{align*}
We thus finally obtain, using the equivariance for vertical vector fields \eqref{Gen.VVF.Equiv}:
\begin{align}
\label{GenGT-X}
\psi_*\bs X_{|p} &=  R_{\gamma(p)*} \bs X_{|p} +   \big[  \gamma(p)\- d\gamma_{|p} \big( \bs X_{|p}\big) \big]^v_{|\psi(p)}, \notag \\[1mm]
			 &= R_{\gamma(p)*} \left(  \bs X_{|p} +  \big[ d\gamma_{|p} \big( \bs X_{|p}\big)  \gamma(p)\-   \big]^v_{|p} \right).
\end{align}
Obviously, iterations of generalised vertical transformations would rely on the same formula: e.g. for $\psi^* \vphi^*=(\varphi \circ \psi)^*$ to which corresponds $\gamma\, (\eta \circ R_\gamma)$ by \eqref{comp}, one has,
\begin{align}
\label{Comp-GenGT-X}
(\varphi \circ \psi)^*\bs X 
			 = R_{\gamma\, (\eta \circ R_\gamma)\, *} \left(  \bs X +  \big[ d[\gamma\, (\eta \circ R_\gamma)] \big( \bs X \big)\,  [\gamma\, (\eta \circ R_\gamma)]\-   \big]^v \right).
\end{align}
The case of $k$-iterations is obvious, though complicated, relying on \eqref{comp-k}.
\medskip

To illustrate, consider the general vertical transformations of a (principal) connection $\omega \in  \C$ and of a tensorial form $\alpha \in \Omega^\bullet_{tens}(P, \rho)$ -- for $\rho:H\rarrow GL(V)$ a representation of $H$. By definition of a connection, $R^*_h \omega_{|ph}=\Ad_{h\-} \omega_{|p}$ and $\omega_{|p}(X^v_{|p})=X \in$ Lie$H$, so one has: 
\begin{align*}
{\omega^\gamma}_{|p}(\bs X_{|p})\defeq \psi^*\omega_{|\psi(p)}(\bs X_{|p})&= \omega_{|\psi(p)}(\psi_*\bs X_{|p}), \\
								  &=  \omega_{|\psi(p)} \left( R_{\gamma(p)*} \bs X_{|p} +   \big[  \gamma(p)\- d\gamma_{|p} \big( \bs X_{|p}\big) \big]^v_{|\psi(p)} \right), \\
			  &= R_{\gamma(p)}^*  \omega_{|\psi(p)} \big( \bs X_{|p} \big)  +   \omega_{|\psi(p)} \left(  \big[  \gamma(p)\- d\gamma_{|p} \big( \bs X_{|p}\big) \big]^v_{|\psi(p)}  \right). \\
			  &= \Ad_{\gamma(p)\-} \omega_{|p}  \big( \bs X_{|p} \big) + \gamma(p)\- d\gamma_{|p} \big( \bs X_{|p}\big).
\end{align*}
The generalised vertical transformation of a connection is thus, 
\begin{align}
\label{GenGT-connection}
\omega^\gamma = \Ad_{\gamma\-} \omega + \gamma\-d\gamma.
\end{align}
It is all but identical to a standard gauge transformation under $\psi \in \Aut_v(P) \sim \gamma \in \H$.
But, doing the same using~\eqref{Comp-GenGT-X}, one obtains the less familiar result for two consecutive general gauge transformations:
\begin{align}
\label{GenGT-connection-2}
(\omega^\eta)^\gamma \defeq \psi^*(\vphi^* \omega) = (\vphi\circ \psi)^*\omega 
 									       &=  \Ad_{[\gamma\, (\eta \circ R_\gamma)]\-} \omega + [\gamma\, (\eta \circ R_\gamma)]\-d[\gamma\, (\eta \circ R_\gamma)], \\
									       &=  \Ad_{ (\eta \circ R_\gamma)\-} \omega^\gamma + (\eta \circ R_\gamma)\-d(\eta \circ R_\gamma). \notag
\end{align}
For $\vphi, \psi \in \Aut_v(P) \sim \eta, \gamma \in \H$ we have $\gamma\, (\eta \circ R_\gamma)=\eta\gamma$, so that we get the standard result 
$(\omega^\eta)^\gamma = \omega^{\eta\gamma}$, expressing the well-known fact that the gauge group $\H$ acts on the right on the space of connections $\C$ -- Hence the fact that the latter can be seen (under proper restrictions) as a principal bundle $\Phi=\C$ with structure group $\H$ (see e.g. \cite{Singer1978, Singer1981}, \cite{Gomes-et-al2018, Francois2021, Francois-et-al2021}).
 
 By definition of a tensorial form $\alpha \in \Omega^\bullet_{tens}(P, \rho)$: $R^*_h \alpha_{|ph}=\rho(h\-) \alpha_{|p}$ and $\alpha_{|p}(X^v_{|p})=0$.
 By the same method, one finds the general vertical transformations,
 \begin{align}
\label{GenGT-tensorial}
\alpha^\gamma &=\rho\big(\gamma\-\big) \alpha, \qquad 
(\alpha^\eta)^\gamma =  \rho\big(\gamma\, (\eta \circ R_\gamma)\big)\- \alpha, 
				    =  \rho \big(\eta \circ R_\gamma \big)\- \alpha^\gamma. 
\end{align}
As a special case, this gives the general vertical transformation of the curvature $\Omega \in \Omega^2_{tens}(P, \Ad)$ of $\omega$. 
Naturally, \eqref{GenGT-tensorial} generalises the well-known gauge transformations of $\alpha$ under $\Aut_v(P)\simeq \H$, 
$(\alpha^\eta)^\gamma = \alpha^{\eta\gamma}$, showing that $\H$ acts on the right on $\Omega^\bullet_{tens}(P, \rho)$ as it does on $\C$.

\subsection{Discussion}
\label{Discussion}

At this point, it should be highlighted that a priori $\omega^\gamma \notin \C$ and $\alpha^\gamma \notin \Omega_{tens}(P, \rho)$. 
To show this, let us first derive the following lemma: the special case of \eqref{GenGT-X} for $\bs X=X^v$ is,  
 \begin{align}
 \label{push-vertic-vect}
 \psi_* X^v_{|p}&= R_{\gamma(p)*}X^v_{|p} + \big[  \gamma(p)\- d\gamma_{|p} \big(  X^v_{|p}\big) \big]^v_{|\psi(p)}, \notag \\
 		       &= \big(\Ad_{\gamma(p)\-} X \big)^v_{|\psi(p)} + \big[  \gamma(p)\- [X^v(\gamma)](p)\big) \big]^v_{|\psi(p)}, \quad \text{using \eqref{Gen.VVF.Equiv},} \notag \\
		       &=\left(\Ad_{\gamma(p)\-} X +  \gamma(p)\- [X^v(\gamma)](p)  \right)^v_{|\psi(p)}.
  \end{align}
  The definition of $\Diff_v(P)\simeq C^\infty(P, H)$ leaves the H-equivariance of $\psi\sim \gamma$ unspecified, so $X^v(\gamma)$ remains a priori unknown.
  Yet, there are two interesting special cases worth emphasazing:
   \begin{align}
   \label{special cases}
 \psi_* X^v_{|p} = \left\{\begin{matrix}  \hspace{-4.2cm} X^v_{|\psi(p)} \ \ \   \text{ for } \psi \in \Aut_v(P)\sim \gamma \in \H \ \text{ (gauge transformation)}, \\[1.5mm]
		       					\ \  0 \ \ \ \ \ \ \ \    \text{ for } \psi(p)=f(p)\defeq pu(p) \text{ with } \gamma=u: P \rarrow H \text{ s.t. } R^*_h u = h\- u \ \text{ (dressing field)}. \end{matrix} \right.
  \end{align}
  The second case involves the \emph{dressing map} $f:P \rarrow P$, $p \mapsto pu(p)$, associated to the \emph{dressing field} $u$, satisfying $f \circ R_h=f$ and thus also $f \circ \psi =f$ for $\psi \in \Diff_v(P)$. It is key to the ``dressing field method" of gauge symmetry reduction, a tool to build gauge-invariants in gauge field theory, see e.g. \cite{Francois-et-al2021, Francois2021, Francois2023-a}. 
  The map $f$ is used to ``\emph{dress}" forms  on $P$, acting via $f^*$, turning them into  \emph{basic} forms (a.k.a ``dressed/composite fields"). 
  
  Indeed, for some $\beta \in \Omega^\bullet(P)$ (horizontality and equivariance unspecified), we define its dressing by $\beta^u\defeq f^*\beta$, satisfying:
     \begin{align}
   \label{generic dressing}
   \beta^u(X^v, \ldots) \defeq&\, f^*\beta\,(X^v \ldots)=\beta (f_*X^v, \ldots)= \beta( 0, \ldots)=0, \notag \\
   R^*_h\, \beta^u \defeq&\,  R^*_h\,f^*\beta = (f\circ R_h)^*\beta = f^*\beta =\beta^u.
     \end{align}
     That is $\beta^u \in \Omega^\bullet_{\text{basic}}(P)$. 
      For example, a dressed connection is  $\omega^u \defeq f^* \omega = \Ad_{u\-} \omega u + u\-du$, while a dressed tensorial form is $\alpha^u =\rho(u\-)\alpha$ -- e.g. the dressed curvature is $\Omega^u= u\-\Omega u = d\omega^u +\sfrac{1}{2}[\omega^u, \omega^u]$.
     Any such dressed form is invariant under $\Diff_v(P)\simeq C^\infty(P, H)$, since 
\begin{align}
\label{inv-dressed-forms}
   (\beta^u)^\gamma \defeq  \psi^* (\beta^u) = \psi^* (f^*\beta) = (f\circ \psi)^* \beta = f^*\beta \defeq \beta^u,
\end{align}
as can also be found via \eqref{GenGT-X} and \eqref{generic dressing}. 
Thus, dressed objects are in particular gauge invariant, i.e. invariant under $\Aut_v(P)\simeq\H$: hence their physical interest, as they may encode physical degrees of freedom (d.o.f.).

Returning to the generic case, \eqref{push-vertic-vect}, for $\omega \in \C$ we have:
 \begin{align}
 \omega^\gamma(X^v)&= \psi^* \omega\, (X^v)=\omega(\psi_*X^v)= \Ad_{\gamma\-} X + \gamma\- X^v(\gamma), \qquad \text{and}\\ 
 R^*_h \omega^\gamma &= R^*_h \psi^* \omega =(\psi \circ R_h)^*\omega. 
 \end{align}
In general then, $\omega^\gamma=\psi^*\omega \notin \C$ for $\psi \in \Diff_v(P) \sim \gamma \in C^\infty(P, H)$. But  per \eqref{special cases},  
   we see the especially significant role played by the groups $\Aut_v(P)\simeq \H$, since only for them do we have  $\psi \circ R_h = R_h \circ \psi$  and $X^v(\gamma)=[\gamma, X]$, so  that $ \omega^\gamma(X^v)=X$ and $R^*_h \omega^\gamma =\Ad_{h\-} \omega^\gamma$. 
   That is,  $\Aut_v(P)\simeq \H$ is the only subgroup of  vertical transformations $\Diff_v(P)\simeq C^\infty(P, H)$ preserving the space of connections $\C$ or $P$.\footnote{ \label{1}This is to be expected as $\Aut(P) \subset \Diff(P)$ is the largest natural transformation group of a bundle $P$, preserving its fibration structure, and $\Aut(P) \cap \Diff_v(P)=\Aut_v(P)$. }
  From above, it is obvious that the second special case in \eqref{special cases} gives  $\omega^u \notin \C$: i.e. a dressed connection is not a connection.
 
 In the same way, for $\alpha \in \Omega^\bullet_{\text{tens}}(P, \rho)$ we have:
  \begin{align}
 \alpha^\gamma(X^v)&= \psi^* \alpha\, (X^v)=\alpha(\psi_*X^v)= 0, \qquad \text{and}\\  
 R^*_h \alpha^\gamma &= R^*_h \psi^* \alpha =(\psi \circ R_h)^*\alpha. 
 \end{align}
 So, if  horizontality is preserved by the action of $\Diff_v(P)\simeq  C^\infty(P, H)$, this is a priori not the case of  $H$-equivariance. Again, only for $\psi \in \Aut_v(P)$ do we get indeed $R^*_h \alpha^\gamma = \rho(h\-)\, \alpha^\gamma$: i.e.  $\Aut_v(P)\simeq \H$ is the only subgroup of  vertical transformations  $\Diff_v(P)\simeq C^\infty(P, H)$ preserving the space of tensorial forms $\Omega^\bullet_{\text{tens}}(P, \rho)$. 
  Again, from above it is obvious that the second special case in \eqref{special cases} gives  $\alpha^u \notin \Omega^\bullet_{\text{tens}}(P, \rho)$.
  
    This shows the importance of being mindful of the equivariance of $\psi \sim \gamma$, if one wants to keep track of the mathematical space in which one evolves: in particular one must keep in mind the clear distinction between (generalised) gauge transformations and dressing operations \eqref{special cases}.
    We stress that, a dressing map $f$ generated by  a dressing field $u$ is not a vertical diffemorphism of $P$. We thus exclude this case of our discussion whenever we mention generalised gauge transformations.   
    
\bigskip
 Connections $\omega \in \C$ can be understood as necessary to obtain  covariant derivatives preserving tensorial forms, i.e. 
$D=d \,+ \rho_*(\omega) :  \Omega^\bullet_{\text{tens}}(P, \rho) \rarrow  \Omega^{\bullet+1}_{\text{tens}}(P, \rho)$ -- which the de Rham derivative $d$ does not ($d\alpha$ fails to be horizontal). 
This implies that $\alpha$ and $D\alpha$ have the same gauge transformations: this is a way to phrase physics' \emph{Gauge Principle}. 
But it also means that $\alpha$ and $D\alpha$ have the same generalised vertical transformations under $\Diff_v(P)\simeq C^\infty(P, H)$: 
\begin{align}
  D\alpha \in   \Omega^{\bullet+1}_{\text{tens}}(P, \rho) \quad \Rightarrow \quad (D\alpha)^\gamma = \rho(\gamma\-)D\alpha.
\end{align}
     Yet, it is also the case that:
 \begin{equation}
 \begin{aligned}
(D\alpha)^\gamma = \psi^* D\alpha = \psi^* \big(d \alpha+ \rho_*(\omega)\alpha \big)
			 			    &=d\psi^*\alpha + \rho_*(\psi^*\omega)\psi^*\alpha, \\
						    &\rdefeq d\alpha^\gamma + \rho_*(\omega^\gamma) \alpha^\gamma.  
\end{aligned}
 \end{equation}
This goes to show that the covariant derivative $D$ is compatible with the action of $\Diff_v(P)\simeq C^\infty(P, H)$, even if under its action it may be that $\omega^\gamma \notin \C$ and $\alpha^\gamma \notin \Omega^\bullet_{tens}(P, \rho)$: In this case, if $D^\gamma \defeq d\, +\rho_*(\omega^\gamma)$ is not a covariant derivative in the geometrical sense of preserving $\Omega^\bullet_{tens}(P, \rho)$, it is in the algebraic sense -- dear to physicists -- of preserving the form of the $\Diff_v(P)\simeq C^\infty(P, H)$ transformation of $\alpha^\gamma$ -- by \eqref{GenGT-connection-2}-\eqref{GenGT-tensorial}. If this transformation is instead a dressing operation, we have in particular that: 
$(D\alpha)^u=D^u \alpha^u= d\alpha^u + \rho_*(\omega^u) \alpha^u$.

\subsection{Infinitesimal vertical transformations}
\label{Finite vertical transformations}

Naturally, the infinitesimal version of the general vertical transformation $\beta^\gamma=\psi^*\beta$ for $\psi \in \Diff_v(P)\sim \gamma \in C^\infty(P, H)$ is given by the Lie derivative along the associated general vertical vector field $X^v \in \diff_v(P) \sim X \in C^\infty(P, \text{Lie}H)$, \eqref{Gen-vert-element}-\eqref{GenVectField}:
\begin{align}
\label{infinit-gen-vertic-trsf}
L_{X^v}\, \beta \defeq \tfrac{d}{d\tau}\, \beta^{\gamma_\tau}\, \big|_{|\tau=0} = \tfrac{d}{d\tau}\, \psi_\tau^*\, \beta \, \big|_{|\tau=0}.
\end{align}
The Lie derivative belongs to the Lie algebra of derivations of $\Omega^\bullet(P)$, it is the case that it is defined by $L_{X^v}=[\iota_{X^v}, d]$ (Cartan's formula). 
Since, $X^v$ can be seen as an element of $\Omega^0(P, VP)$, $L_{X^v}$ is to be understood as a special case of  the Nijenhuis-Lie derivative along the vertical vector-valued $0$-form $X^v$.
Naturally then, by   \eqref{NL-FN-der-ext-bracket}, the commutator of Nijenhuis-Lie derivatives  involves the Frölicher-Nijehnhuis brachet on $\Omega^0(P, VP)$, i.e. the extended bracket $\{\, ,\, \}$ on $C^\infty(P, \text{Lie}H)$ \eqref{ext-bracket}-\eqref{FN-bracket}:
\begin{align}
\label{commutator-inf-gen-vert-trsf}
[L_{X^v}, L_{Y^v}]\, \beta = L_{[X^v, Y^v]_{\text{\tiny{FN}}}}\, \beta =  L_{\{X, Y\}^v}\, \beta.  
\end{align}

To illustrate, let us have the $ \diff_v(P)\simeq \Omega^0(P, VP) \sim \C^\infty(P, \text{Lie}H)$ transformations of $\omega\in \C$ and $\alpha \in \Omega^\bullet_{tens}(P, \rho)$. 
For~a~connection, using \eqref{GenGT-connection}, we get:
\begin{align}
\label{Lie-vert-trsf-connection}
L_{X^v} \omega \defeq  \tfrac{d}{d\tau}\, \psi_\tau^*\, \omega \, \big|_{|\tau=0} =  \tfrac{d}{d\tau}\,   \Ad_{\gamma\-_\tau} \omega +  \gamma\-_\tau d \gamma_\tau\, \big|_{|\tau=0} = -\ad_X \omega + dX = DX. 
\end{align}
This is cross-checked via Cartan's formula, using Cartan's structure equation for the curvature, and the latter's tensoriality: For $\bs X \in \Gamma(TP)$, we have in general   
\begin{align} 
L_{\bs X} \omega = \iota_{\bs X} d \omega + d(\iota_{\bs X} \omega) =  \iota_{\bs X} \big(\Omega - \tfrac{1}{2}[\omega, \omega]\big) + d\big(\omega(\bs X)\big)=   \iota_{\bs X} \Omega + d\big(\omega(\bs X)\big) + [\omega, \omega(\bs X)].
\end{align} 
Thus, for $\bs X =X^v $ we get $L_{X^v} \omega = dX +[\omega, X]=DX$.
We must remain mindful that  if  $X \notin \Omega^0_{\text{tens}}(P, \Ad)=$ Lie$\H$ then  $DX=dX + [\omega, X] \notin \Omega^1_{\text{tens}}(P, \Ad)$: only for $X^v \in \aut_v(P) \sim X \in$ Lie$\H$ does $DX$ answers the geometric definition of the covariant derivative on tensorial forms.  
This echoes the observation made earlier that  $\omega^\gamma=\psi^*\omega \notin \C$ for a generic $\psi \in \Diff_v(P)\sim \gamma \in C^\infty(P, H)$.

As $\C$ is an affine space modeled on $\Omega^1_{\text{tens}}(P, \Ad)$,  for $\omega', \omega \in C$ it must be that $\omega'-\omega \in \Omega^1_{\text{tens}}(P, \Ad)$. So, in particular the linear action of a transformation group of $\C$ would result in an element of $\Omega^1_{\text{tens}}(P, \Ad)$. As a matter of fact, $\Aut(P)$ is the maximal such group: Indeed, for $\psi \in \Aut(P)$ it is the case that,
\begin{align}
\label{push-vertic-by-Aut}
\psi_*X(p)^v_{|p} \defeq \tfrac{d}{d\tau}\,  \psi\big( \phi_\tau(p) \big)  \,\big|_{\tau=0} 
				= \tfrac{d}{d\tau}\,  \psi\big( R_{\gamma_\tau(p)}\, p \big)  \,\big|_{\tau=0} 
			       = \tfrac{d}{d\tau}\,   R_{\gamma_\tau(p)} \psi(p)  \,\big|_{\tau=0} \rdefeq X(p)^v_{|\psi(p)}. 
\end{align} 
So we have first $\psi^*\omega_{|p}(X^v_{|p})=\omega_{|\psi(p)}(X^v_{|\psi(p)})=X$, and second 
$R^*_h \psi^*\omega = \psi^* R^*_h \omega = \psi^* \Ad_{h\-} \omega= \Ad_{h\-} \psi^*\omega$: i.e. $\psi^*\omega \in \C$. 
It follows indeed that $L_{\bs X} \omega  \in \Omega^1_{\text{tens}}(P, \Ad)$ for $\bs X \in \aut(P)$,  thus also for the special case $\bs X=X^v \in \aut_v(P)$ with explicit result $ L_{X^v} \omega =DX \in \Omega^1_{\text{tens}}(P, \Ad)$. 

Applying a second Nijenhuis-Lie derivative we get, using the fact that $[L_{\bs X}, d]=0$:
\begin{align*}
L_{Y^v}L_{X^v} \omega = d(L_{Y^v}X) + [L_{Y^v} \omega, X] + [\omega, L_{Y^v} X] &= d\big(  Y^v(X)\big) + [DY, X] + [\omega,  Y^v(X)], \\
																 &= D\big( Y^v(X) \big) + [dY, X] + \big[[\omega, Y], X\big]
\end{align*}
From this follows that, 
\begin{align}
\label{Com-omega}
[L_{X^v}, L_{Y^v}] \omega &=  D\big( X^v(Y) \big) + [dX, Y] + \big[[\omega, X], Y\big] - D\big( Y^v(X) \big) - [dY, X] - \big[[\omega, Y], X\big],  \notag\\
					 &= D\big( X^v(Y) \big)  - D\big( Y^v(X) \big) + d\big( [X, Y] \big)  +  \big[[\omega, X], Y\big]   - \big[[\omega, Y], X\big], \notag\\
					 &= D\big( X^v(Y) \big)  - D\big( Y^v(X) \big) + d\big( [X, Y] \big)  +  \big[\omega, [X, Y] \big], \notag \\
					 &= D\big( \{X,Y\} \big) = L_{\{X, Y\}^v} \omega = L_{[X^v, Y^v]_{ \text{ \tiny{FN} }}} \omega, 
\end{align}
by Jacobi identity in Lie$H$ from the second to third line, and by definition of the extended/FN bracket \eqref{ext-bracket}-\eqref{FN-bracket-field-dep-diff=BT-bracket} from the third to the last. This result is as expected from \eqref{commutator-inf-gen-vert-trsf}.

Similarly for a $\alpha \in \Omega^\bullet_{tens}(P, \rho)$ we have, 
\begin{align}
\label{Lie-vert-trsf-alpha}
L_{X^v} \alpha \defeq  \tfrac{d}{d\tau}\, \psi_\tau^*\, \alpha \, \big|_{|\tau=0} =  \tfrac{d}{d\tau}\,   \rho(\gamma\-_\tau) \alpha  \,\big|_{|\tau=0} = -\rho_*(X) \alpha. 
\end{align}
This can be cross-checked via Cartan's formula: For $\bs X \in \Gamma(TP)$, we have in general   
\begin{align} 
L_{\bs X} \alpha &= \iota_{\bs X} d \alpha + d(\iota_{\bs X} \alpha) =  \iota_{\bs X} \big(D\alpha - \rho_*(\omega) \alpha+ d(\iota_{\bs X} \alpha)
												=   \iota_{\bs X} D\alpha  - \rho_*( \iota_{\bs X} \omega) \alpha +  \rho_*(\omega)  \iota_{\bs X} \alpha+ d(\iota_{\bs X} \alpha) \notag \\
												&=  \iota_{\bs X} D\alpha  + D(\iota_{\bs X} \alpha) - \rho_*( \iota_{\bs X} \omega) \alpha. 
\end{align} 
Thus, for $\bs X =X^v $ we get $L_{X^v} \alpha = -\rho_*(X)\alpha$. 
Here again, we notice that since generically $X \notin \Omega^0_{\text{tens}}(P, \Ad)$, generically $L_{X^v} \alpha \notin \Omega^\bullet_{tens}(P, \rho)$. Which echoes the earlier observation that $\alpha^\gamma=\psi^*\alpha \notin \Omega^\bullet_{tens}(P, \rho)$ for a generic $\psi \in \Diff_v(P)\sim \gamma \in C^\infty(P, H)$. Upon applying the Nijenhuis-Lie derivative again we get, 
\begin{align*}
L_{Y^v}L_{X^v} \alpha = -\rho_*(L_{Y^v}X) \alpha - \rho_*(X) L_{Y^v} \alpha = -\rho_*\big( Y^v(X) \big) \alpha +  \rho_*(X)  \rho_*(Y) \alpha. 
\end{align*}
Then it follows that, 
\begin{align}
\label{Com-alpha}
[L_{X^v}, L_{Y^v}] \alpha &= -\rho_*\big( X^v(Y) \big) \alpha +  \rho_*(Y)  \rho_*(X) \alpha + \rho_*\big( Y^v(X) \big) \alpha -  \rho_*(X)  \rho_*(Y) \alpha, \notag\\
					    &= -\rho_*\big( [X,Y] + X^v(Y) - Y^v(X)  \big) \alpha, \notag \\
					    &= -\rho_*\big(  \{ X, Y \}\big)\alpha =L_{\{X, Y\}^v} \alpha = L_{[X^v, Y^v]_{ \text{ \tiny{FN} }}} \alpha.
\end{align}
Which is as expected from \eqref{commutator-inf-gen-vert-trsf}.
\medskip

In the next and final section, we show how the global structures on $P$ exposed up to now descend as local structures on the base space $M$. These may be noteworthy for gauge field theory, as vertical transformations generalise the usual (active or passive) gauge transformations. A gauge argument still applies to them. 
\medskip

\clearpage

\section{Local structure}
\label{Local structure}

We first  summarize the standard local structure, and extend it to general vertical transformations in the next section.

\subsection{Gluings and local active gauge transformations}
\label{Gluings and local active gauge transformations}

Given a open subset $U \subset M$ and a local section $\s :U \rarrow P_{|U}$, a local representative of  $\beta \in \Omega^\bullet(P)$ is $b \defeq \s^*\beta \in \Omega^\bullet(U)$. 
Any other local section is $s'=sg$ with $g:U \rarrow H$  a transition function of the bundle (encoding its topology, from a local viewpoint), so another local representative is $b' \defeq {\s'}^* \beta =b^g$. The notation $b'=b^g$ is meant to indicate that $b'$ can be seen as obtained from $b$ by a right action transformation by a transition function $g$. 
For yet another local section $s''=s'g' =sgg'$, we have that $b''=(b')^{g'}=b^{gg'}$. These constitute the \emph{gluing relations} of local representatives\footnote{The terminology stems from considering $s$,  $s'$ and $s''$ as local sections over overlapping opens $U$, $U'$ and $U''$ respectively, related by the transition functions $g$ and $g'$. Then the gluing relations reflects the fact that the local representatives on $M$ , the $b$'s, comes from the same global object $\beta$ on $P$ -- which can be reconstructed from its local representatives.} and are known in gauge field theory as \emph{passive gauge transformations}.

To illustrate, for  $\beta =\omega \in \C$, we define $A \defeq \s^*\omega\, \in \A$ (the gauge potential) and we have the well known result $A'=\Ad_{g\-} A+g\-dg\rdefeq A^g$, and $A''=\Ad_{{g'}\-} A' +{g'}\-dg' = A^{gg'}$.
In the case of $\beta=\alpha \in \Omega^\bullet_\text{tens}(P, \rho)$, one has $a\defeq \s^*\alpha\, \in \Omega^\bullet_\text{tens}(U, \rho)$ and it is well-known that $a'=\rho(g\-)\,a\rdefeq a^g$, and $a''=\rho({g'}\-)\,a'=\rho\big((gg')\-\big)\,a=a^{gg'}$. 
The local representative of the curvature $\Omega \in \Omega^2_\text{tens}(P, \Ad) $ of $\omega$ is $F\defeq \s^*\Omega$. As a special case of the above, we have then $F'=\Ad_{g\-} F\rdefeq F^g$ and $F''=\Ad_{{g'}\-} F' = F^{gg'}$. 
A matter field would be some $\alpha =\phi \in \Omega^0_\text{tens}(P, \rho)$ with local representative $a=\upphi$. Its covariant derivative $D^\omega\phi =d\phi +\rho_*(\omega)\phi$ has local representative $D^A\upphi=d\upphi + \rho_*(A)\upphi$, which represents the minimal coupling to the gauge potential. 
It is well known that $(D^A\upphi)^g=D^{A^g} \upphi^g=\rho(g\-)D^A \upphi$, which expresses the requirement of the (passive) \emph{gauge principle}.
\medskip

Of course, the group $\Aut_v(P)$ cannot act on local representatives, but there is naturally such a thing as the local representative  of the gauge transformed $\beta^\gamma \defeq \psi^*\beta$, for  $\gamma \in  \H\simeq \psi \in \Aut_v(P)$. 

Given the defining equivariance of elements of $\H$, making them tensorial 0-forms for the conjugate action of $H$, the $\Aut_v(P)\simeq\H$-transformation of $\eta\in \H$ is -- as a special case of the one discussed below \eqref{GenGT-tensorial} -- $\eta^\gamma\defeq \psi^*\eta=\gamma\- \eta \gamma$. This relation defines the action of gauge group $\H$ on itself. 
 Defining $\upgamma \defeq \s^*\gamma$ and $\upeta \defeq \s^*\eta$, we have $\upeta^\upgamma \defeq \s^*( \eta^\gamma)=\upgamma\- \upeta \upgamma$. Therefore, the local representative of $\H$, the \emph{local gauge group}, is defined as:
  $\H_{\text{\tiny{loc}}} \defeq \{ \upeta, \upgamma: U \rarrow H\, |\, \upeta^\upgamma =\upgamma\- \upgamma \upgamma\}$.
 
  We have therefore the local representative of a gauge transformed form: $b^{\upgamma}\defeq \s^*(\beta^\gamma)$. 
  Given that $(\beta^\eta)^\gamma =\beta^{\eta\gamma}$, it is the case that $(b^{\upeta})^\upgamma =b^{\upeta\upgamma}$.
 This secures the consistency of a heurisitic ``field theoretic rule": Considering $\upgamma$ as acting on $b^{\upeta}$ seen as a concatenation of the fields $b$ and $\upeta$, one may write $(b^{\upeta})^\upgamma = (b^\upgamma)^{\upeta^\upgamma}$, using then their respective $\H_{\text{\tiny{loc}}}$-transformations, one writes the concatenation as $(b^\upgamma)^{\upeta^\upgamma} =(b^\upgamma)^{\upgamma\-\upeta\upgamma}=b^{\upgamma\, \upgamma\-\upeta\upgamma} =b^{\upeta\upgamma}$. 
 
 The relation $(b^{\upeta})^\upgamma =b^{\upeta\upgamma}$ expresses the fact that the action of the local gauge group $\H_{\text{\tiny{loc}}}$ on local representatives -- the fields of physics -- is a right action. 
 Hence, the field space $\Phi=\{b\}$ can be seen (under proper restrictions) as an infinite dimensional bundle with structure group 
 $\H_{\text{\tiny{loc}}}$. A fact rarely stressed, or fully exploited, in the covariant phase space literature -- see e.g. \cite{Gomes-et-al2018, Gomes-Riello2021, Francois-et-al2021, Francois2021, Francois2023-a}.

 To illustrate as above, we have that $A^\upgamma \defeq \s^*(\omega^\gamma) = \Ad_{\upgamma\-} A+\upgamma\-d\upgamma$. And $(A^{\upeta})^\upgamma =A^{\upeta\upgamma}$, as can be checked from using the previous definition and $(A^{\upeta})^\upgamma = (A^\upgamma)^{\upeta^\upgamma}$. 
 Similarly for a tensorial form we have $a^\upgamma \defeq \s^*(\alpha^\gamma) = \rho(\upgamma\-)\, a$, from which one checks that $(a^{\upeta})^\upgamma = a^{\upeta\upgamma}$ via $(a^{\upeta})^\upgamma =(a^{\upgamma})^{\upeta^\upgamma}$. 
 As a special cases, $F^\upgamma \defeq \s^*(\Omega^\gamma) = \Ad_{\upgamma\-} F$, and $\upphi^\upgamma \defeq \s^*(\phi^\gamma) = \rho(\upgamma\-)\, \upphi$ while its covariant derivative is 
 $(D^A\upphi)^\upgamma \defeq \s^*\big((D^\omega\phi)^\gamma\big) = \rho(\upgamma\-)\, D^A\upphi$. Here, the local covariant derivative operator can be \emph{defined} as the expression $D\defeq d\ + \rho_*(A)$. 
 It is then easily checked that the previous result is also recovered from the field theoretic rule:  $(D^A\upphi)^\upgamma =D^{A^\upgamma }\upphi^\upgamma = \rho(\upgamma\-)\, D^A\upphi$, which is yet another expression of the gauge principle. 
 \medskip
 
Notice that, from the standpoint of gauge field theory, there is no way to distinguish between local gluings $b^g$ and the action $b^\upgamma$ of 
$\H_{\text{\tiny{loc}}}$, i.e. between passive gauge transformations and local active gauge transformations. The ``gauge principle" therefore actually encapsulates two gauge principles: a passive and an active one.
Yet, the conceptual meaning of each is quite different, as different as coordinate changes (a.k.a. passive diffeomorphisms) and (active) diffeomorphisms $\Diff(M)$ are in GR. 

A theory being given by a Lagrangian $L$, invariance under gluings, $L(b)=L(b^g)$,  mean that a gauge field theory is only sensitive to the intrinsic geometry of the bundle $P$, and to global object $\beta$ living in it. Invariance under $\H_{\text{\tiny{loc}}}$, 
$L(b)=L(b^\upgamma)$, means that a gauge field theory is sensitive only to the geometry of the $\Aut_v(P)$-class of $P$, i.e. to the $\H$-orbits of global objects. 
In the same way that the combination of the hole argument and the point-coincidence argument \cite{Giovanelli2021} clarified that  $\Diff(M)$-invariance in GR encodes the \emph{relational} nature of spacetime, and of general relativistic physics more generally, arguably a combination of an ``internal hole argument" and ``internal point-coincidence argument" suggests that $\H_{\text{\tiny{loc}}}$-invariance encodes the  \emph{relational} character of the enriched spacetime, represented by a (class of) principal bundle, and of gauge field physics more generally. We will elaborate on this in a forthcoming work. 

 \subsection{Linear versions}
\label{Linear versions}

All of the above naturally have linear versions. Consider a transition function $g_\tau :U \rarrow H$, so that $\s'_\tau=\s g_\tau$,  s.t. $g_{\tau=0}=\id_{\text{\tiny{H}}}$.
The object $\lambda =\tfrac{d}{d\tau} g_\tau\, \big|_{\tau=0}: U \rarrow$ Lie$H$ is an infinitesimal transition function of the bundle $P$. 
We~define $\delta_{\!\lambda\,} b \defeq \tfrac{d}{d\tau}\, b^{g_\tau}\, \big|_{\tau=0}$: It is the limit of the difference between the local representatives obtained via $\s'_\tau$ and $\s$ respectively. Such infinitesimal gluings can be called  an infinitesimal passive gauge transformations.

A commutator $[\delta_{\!\lambda}, \delta_{\!\lambda'}] b=\delta_{[\lambda,\, \lambda']_{\text{\tiny{Lie$H$}}}} b$ arises from the definition 
$[\delta_{\!\lambda}, \delta_{\!\lambda'}] b \defeq \tfrac{d}{ds}\tfrac{d}{d\tau}\, b^{g_\tau g'_sg\-_\tau {g'}\-_s}\, \big|_{\tau=0}\big|_{s=0}$  -- i.e. the  commutator in $H$. It is of course recovered when one has concrete expressions for the result of $\delta_{\!\lambda\,} b$, and considering $\delta_{\!\lambda}$ as an even graded derivation on $\Omega^\bullet(U)$ s.t. $[\delta_{\!\lambda\,}, d]=0$, and whose action on $b$'s is defined by such expressions. 
The~commutator is then of course $[\delta_{\!\lambda}, \delta_{\!\lambda'}] =\delta_{\!\lambda} \delta_{\!\lambda'} - \delta_{\!\lambda} \delta_{\!\lambda'}$.
 Let us illustrate.

From above, we obtain well-known relations, for $A \in \A$ and $a\in \Omega^\bullet_\text{tens}(U, \rho)$: First, 
$\delta_{\!\lambda\,} A \defeq \tfrac{d}{d\tau}\, A^{g_\tau}\, \big|_{\tau=0} = d\lambda - \ad_{\lambda} A = d\lambda + [A, \lambda]=D^A\!\lambda$. 
Then, $\delta_{\!\lambda\,} a \defeq \tfrac{d}{d\tau}\, a^{g_\tau}\, \big|_{\tau=0} = -\rho_*(\lambda)\,a$.
As special case we have, $\delta_{\!\lambda\,} F =- \ad_{\lambda} F =[F, \lambda]$, $\delta_{\!\lambda\,} \upphi = -\rho_*(\lambda) \upphi$ and 
$\delta_{\!\lambda\,} D^A\upphi = -\rho_*(\lambda) D^A\upphi$. The latter result is recovered, seeing $\delta_{\!\lambda}$ as an even derivation, via $\delta_{\!\lambda\,} D^A\upphi = d \delta_{\!\lambda\,} \upphi +[\delta_{\!\lambda\,}A, \upphi] - [A, \delta_{\!\lambda\,}\upphi]$. 
In the same algebraic way, it~it~easily verified on $A$ and $a$ that 
$[\delta_{\!\lambda}, \delta_{\!\lambda'}] A =\delta_{\!\lambda} \delta_{\!\lambda'}A  - \delta_{\!\lambda} \delta_{\!\lambda'} A = D^A\big([\lambda,\, \lambda']_{\text{\tiny{Lie$H$}}}\big) = \delta_{[\lambda, \,\lambda']_{\text{\tiny{Lie$H$}}}} A$, 
and 
that $[\delta_{\!\lambda}, \delta_{\!\lambda'}] a =\delta_{\!\lambda} \delta_{\!\lambda'} a- \delta_{\!\lambda} \delta_{\!\lambda'}a =-\rho_*\big([\lambda, \lambda']_{\text{\tiny{Lie$H$}}}\big)\,a = \delta_{[\lambda,\, \lambda']_{\text{\tiny{Lie$H$}}}} a$. 
These computations are familiar in gauge field theory.
\medskip

Given the infinitesimal equivariance of elements of Lie$\H$ (see below \eqref{Id-ext-bracket}), making them tensorial 0-form for the $\Ad$-representation of $H$, the $\Aut_v(P)\simeq\H$-transformation of $Y\in$ Lie$H$ is $Y^\gamma\defeq\psi^* Y=\Ad_{\gamma\-} X= \gamma\- X \gamma $. The~corresponding  $\aut_v(P)\simeq$ Lie$\H$-transformation is $ L_{X^v} Y =X^v(Y) =-\ad_X Y=[Y, X]_{\text{\tiny{Lie$H$}}}$.  
Thus, defining $\xi\defeq \s^*X$ and $ \zeta \defeq \s^*Y$, we have locally: $\zeta^\upgamma = \Ad_{\upgamma\-} \zeta$ and the corresponding linearisation $\delta_\xi \zeta \defeq \s^*( L_{X^v} Y) = [ \zeta, \xi]_{\text{\tiny{Lie$H$}}}$. 
We~define the Lie algebra of the local gauge group as Lie$\H_{\text{\tiny{loc}}}\defeq \{\xi, \zeta :U \rarrow \text{Lie}H\ | \ \delta_\xi \zeta= [ \zeta, \xi]_{\text{\tiny{Lie$H$}}} \}$. 

Local active infinitesimal gauge transformations of $b$ are defined as $\delta_\xi b \defeq \s^*(L_{X^v} \beta)$. Here,  $\delta_\xi$ is immediately seen as an even derivation on $\Omega^\bullet(U)$ s.t. $[\delta_\xi, d]=0$ arising from the Lie derivative on $P$ along $X^v \in \aut_v(P)$. 
So,~naturally, a Lie bracket arises by
 $[\delta_\xi, \delta_\zeta] b \defeq \s^*\big([L_{X^v}, L_{Y^v}]\,\beta \big) = \s^*\big( L_{(-[X, Y]_{\text{\tiny{Lie$H$}}})^v} \,\beta \big) = - \delta_{ [\xi, \,\zeta]_{\text{\tiny{Lie$H$}}} } b$. See again below \eqref{Id-ext-bracket}. 
 This is of course cross-checked by writing the bracket as a commutator $[\delta_\xi, \delta_\zeta] =\delta_\xi \delta_\zeta - \delta_\zeta \delta_\xi $,  from explicit expressions for $\delta_\xi b$ seen as defining the action of $\delta_\xi$ on the $b$'s
and by using the action $\delta_\xi \zeta = [\zeta, \xi]_{\text{\tiny{Lie$H$}}}$  of Lie$\H$ on its elements.
  Let us illustrate. 

For $A\in \A$ and  $a\in \Omega^\bullet_\text{tens}(U, \rho)$, we obtain: 
 $\delta_\xi A \defeq \s^*(L_{X^v} \omega) = D^A\xi$ and  $\delta_\xi a \defeq \s^*(L_{X^v} \alpha) = -\rho_*(\xi)\, a$. 
 As~special case we have, $\delta_\xi F =- \ad_{\xi} F =[F, \xi]$, $\delta_\xi \upphi = -\rho_*(\xi)\, \upphi$ and 
$\delta_\xi\, D^A\upphi = -\rho_*(\xi) \,D^A\upphi$.
This last result is recovered algebraically by $\delta_\xi\, D^A\upphi = d \delta_\xi \upphi +[\delta_\xi A, \upphi] - [A, \delta_\xi \upphi]$. 
In the same way, one easily shows:
$[\delta_\xi, \delta_\zeta] A =\delta_\xi \delta_\zeta A - \delta_\zeta \delta_\xi A = -D^A\big([\xi, \zeta]_{\text{\tiny{Lie$H$}}}\big) = -\delta_{[\xi,\, \zeta]_{\text{\tiny{Lie$H$}}}} A$, 
and 
that $[\delta_\xi, \delta_\zeta] a =\delta_\xi \delta_\zeta a- \delta_\zeta \delta_\xi a =\rho_*\big([\xi, \zeta]_{\text{\tiny{Lie$H$}}}\big)\, a=  -\delta_{[\xi,\, \zeta]_{\text{\tiny{Lie$H$}}}} a$. 
\medskip
 
  We observe again that, from a gauge field theoretic perspective, Lie$\H_{\text{\tiny{loc}}}$-transformations are indistinguishable from infinitesimal gluings (apart from a sign difference in their commutators). 
 Both can be encapsulated via the BRST framework \cite{Bertlmann}:
 There, one extends $\Omega^\bullet(U)$ to the bigraded complex $\Omega^\bullet(U, \rho)\otimes \wedge^\bullet c$, where $c$ is the odd degree (Grassmann) Lie$H$-valued  ghost field, place-holder for the parameters $\lambda$ or $\xi \in$ Lie$\H_{\text{\tiny{Loc}}}$. The BRST differential $s$, s.t. $s^2=0$  and $sd=-ds$, is introduced,  playing the role of $\delta_{\!\lambda\,}$ or $\delta_\xi$. The action of $s$ on $b=\{A, a\}$ and $c$ is by definition:
 $sA\defeq-Dc=-dc -Ac-cA$ (a bigraded bracket is understood in $Dc$), $sa \defeq-\rho_*(c)\,\phi$, and $sc\defeq\sfrac{1}{2}[c, c]_{\text{\tiny{Lie$H$}}}$. 
 The~third relation enforces $s^2=0$ on all fields, while the first two  reproduces infinitesimal  gauge transformations.
 This algebraic treatment, efficient as it is, erases the key conceptual difference between passive and active local gauge transformations, i.e. between gluings and Lie$\H_{\text{\tiny{Loc}}}$. 
 As we show below, the  local version of (iterated) general vertical transformations can be distinguished from local gluings. 
 

\subsection{Generalised local active gauge transformations}
\label{Generalised local active gauge transformations}

Like $\Aut_v(P)\simeq \H$, the group $\Diff_v(P)\simeq C^\infty(P, H)$ cannot act on local representatives $b$ of forms $\beta$ on $P$. Yet, one can define the local representatives of $\beta^\gamma\defeq \psi^*\beta$ for $\gamma \in C^\infty(P, H) \sim \psi \in \Diff_v(P)$. 
As the equivariance of elements in $C^\infty(P, H)$ are left unspecified, so are both their $\Aut_v(P)\simeq \H$ and $\Diff_v(P)$ transformations. We thus simply use the generic notation $\eta^\gamma \defeq \psi^*\eta = \eta \circ R_\gamma$, for  $\gamma, \eta \in C^\infty(P, H)$. 
By \eqref{comp}  the composition law in $\Diff_v(P)$ is thus represented by the element $\gamma\, \eta^\gamma \in C^\infty(P, H)$.

We define $\upgamma \defeq \s^*\gamma$, and  $C^\infty(U, H)\defeq \{\upgamma: U\rarrow H \}$ is the local version of the group of vertical transformations, so we may call it  the group of  \emph{generalised local active gauge transformations}. 
Its action on local representatives is given by definition as $b^\upeta \defeq \s^*(\beta^\eta)$. By \eqref{GenGT-tensorial}, the iteration law is given by  $(b^\upeta)^\upgamma \defeq \s^*\big( (\beta^\eta)^\gamma \big) = \s^*\big( \beta^{\gamma\, \eta^\gamma}\big) = b^{\upgamma\, \upeta^\upgamma}\!$. 
 This again secures the consistency of ``field theoretic rule": Considering $\upgamma$ as acting on $b^{\upeta}$ as a concatenation of the fields $b$ and $\upeta$, one may write $(b^{\upeta})^\upgamma = (b^\upgamma)^{\upeta^\upgamma}$, which results indeed in the concatenation $b^{\upgamma\, \upeta^\upgamma}\!$. 

For example, by \eqref{GenGT-connection}-\eqref{GenGT-connection-2} and \eqref{GenGT-tensorial}, the generalised gauge transformations of the gauge potential $A\in \A$ and $a \in \Omega^\bullet_\text{tens}(U, \rho)$ are:
\begin{equation}
\label{Gen-GT-local}
\begin{alignedat}{2}
A^\upeta &= \Ad_{\upeta\-} A + \upeta\-d\upeta \quad &&\text{and}   \quad
(A^\upeta)^\upgamma = \Ad_{(\upgamma\, \upeta^\upgamma)\-} A + (\upgamma\, \upeta^\upgamma)\-d(\upgamma\, \upeta^\upgamma), \\[.5mm]
a^\upeta &= \rho(\upeta)\- a \quad &&\text{and}   \quad
(a^\upeta)^\upgamma = \rho(\upgamma\, \upeta^\upgamma)\- a .
\end{alignedat}
\end{equation}
The second line gives in particular the generalised transformations of the field strength $F$, matter field $\upphi$ and its covariant derivative $D^A\upphi$. The latter is also found via the field-theoretic computation: $(D^A\upphi)^\eta = D^{A^\upeta} \upphi^\upeta$, idem for iterated transformations. 
This is noteworthy:  It shows that the operator $D\defeq d\ +\rho_*(A)$  preserves
the covariance of $a$, or $\phi$, and deserve the name ``covariant derivative",  even under the action $C^\infty(U, H)$, extending that of 
$\H_{\text{\tiny{loc}}}$. 
And this, as discussed at the end of section \ref{Discussion}, despite the fact that $A^\upeta$ is not the local representative of a connection, since  $\omega^\eta \notin \C$, and that $a^\upeta$ is not the local representative of a tensorial form, since $ \alpha^\eta \notin \Omega^\bullet_\text{tens}(P, \rho)$.
This shows that a Gauge Principle, or Gauge Argument, applies still to generalised gauge transformations. 

A Lagrangian invariant under gluings, $L(b^g)=L(b)$, will then not only be invariant under $\H_{\text{\tiny{loc}}}$ as previously mentioned, but also under $C^\infty(U, H)$: $L\big(b^{\upgamma\, \upeta^\upgamma}\big)=L(b^\upeta)=L(b)$. Gauge field theories  thus naturally enjoys a larger group of gauge symmetries, arising from $\Diff_v(P)$.
\medskip

Let us  consider the infinitesimal counterpart of the above.  The Lie algebra of generalised gauge transformations is the local version of $C^\infty(P, \text{Lie}H)$ equipped with the extended bracket 
 \eqref{ext-bracket}. 
 Given that the infinitesimal equivariance of elements of $C^\infty(P, \text{Lie}H)$ is left unspecified, so is their their $\Aut_v(P)\simeq \H$ and $\Diff_v(P)$ transformations. 
 We~thus have, for the action of $\gamma \in C^\infty(P, H)$ on  $Y \in C^\infty(P, \text{Lie}H)$, the generic notation $Y^\gamma\defeq \psi^* Y$. Correspondingly, the action of $X \in C^\infty(P, H)$ is noted $L_{X^v} Y=X^v(Y)$. 
 Thus, defining $\xi\defeq \s^*X$ and $ \zeta \defeq \s^*Y$, we have locally: $\zeta^\upgamma \defeq s^*(Y^\gamma)$ and the corresponding linearisation $\delta_{\xi\,} \zeta \defeq \s^*( L_{X^v} Y)$. The Lie algebra of infinitesimal local generalised gauge transformations is then
 $C^\infty(U, \text{Lie}H)$ equipped with the local  Frölicher-Nijenhuis bracket  \eqref{ext-bracket}-\eqref{FN-bracket-field-dep-diff=BT-bracket}:
 \begin{equation}
\label{ext-bracket-local}
\begin{aligned}
\big\{ \xi, \zeta \big\} = [\xi,  \zeta]_{\text{\tiny{Lie$H$}}} +  \delta_{\xi\,}\zeta -  \delta_{\zeta\,} \xi,
\end{aligned}
\end{equation}
 We may call this a generalised gauge algebra. 
 
Local infinitesimal generalised gauge transformations  of $b$ are defined by $\delta_\xi b \defeq \s^*(L_{X^v} \beta)$, given  \eqref{infinit-gen-vertic-trsf}, where $\delta_\xi$ is immediately seen as an even derivation on $\Omega^\bullet(U)$, s.t. $[\delta_\xi, d]=0$, arising from the Nijenhuis-Lie derivative on $P$ along $X^v \in \diff_v(P)$. 
 Naturally, a Lie bracket for $\delta_\xi$'s arises via $[\delta_\xi, \delta_\zeta]\, b \defeq \s^*\big([L_{X^v}, L_{Y^v}]\,\beta \big)$, i.e. from the commutator of the Nijenhuis-Lie derivatives  \eqref{Id-ext-bracket}, which by \eqref{commutator-inf-gen-vert-trsf} gives:
 \begin{equation}
\label{Gen-gauge-alg}
\begin{aligned}
[\delta_\xi, \delta_\zeta]\, b &= \delta_{\{ \xi,\, \zeta \}\,} b. 
\end{aligned}
\end{equation}
Local fields are  representations for the generalised gauge algebra.  
This is obtained algebraically too, writing the commutator $[\delta_\xi, \delta_\zeta] =\delta_\xi \delta_\zeta - \delta_\zeta \delta_\xi$, and using  explicit expressions for $\delta_\xi b$ seen as defining the action of $\delta_\xi$ on the $b$'s,
as well as the action $\delta_\xi \zeta$  of $C^\infty(U, \text{Lie}H)$ on its elements.

The illustration with  $A\in \A$ and  $a\in \Omega^\bullet_\text{tens}(U, \rho)$ is formally as before: we obtain from \eqref{Lie-vert-trsf-connection}-\eqref{Lie-vert-trsf-alpha},
\begin{align}
 \delta_\xi A \defeq \s^*(L_{X^v} \omega) = D^A\xi, \quad \text{and} \quad   \delta_\xi a \defeq \s^*(L_{X^v} \alpha) = -\rho_*(\xi)\, a. 
 \end{align}
 As~special case of the second equation we have, $\delta_\xi F =- \ad_{\xi} F =[F, \xi]$, $\delta_\xi \upphi = -\rho_*(\xi)\, \upphi$ and 
$\delta_\xi\, D^A\upphi = -\rho_*(\xi) \,D^A\upphi$.
The last result being recovered algebraically by $\delta_\xi\, D^A\upphi = d \delta_\xi \upphi +[\delta_\xi A, \upphi] - [A, \delta_\xi \upphi]$. 
In the same algebraic way, one checks the commutator on $a$:
 \begin{equation}
\begin{aligned}
[\delta_\xi, \delta_\zeta]\, a   & = - \delta_\xi \,\rho_*(\zeta)\, a +  \delta_\zeta\, \rho_*(\xi)\, a, \\
					 & = -  \rho_*(\delta_{\xi\,} \zeta)\, a + \rho_*(\zeta) \rho_*(\xi)\,a  \ 
					       + \  \rho_*(\delta_{\zeta\,} \xi)\, a - \rho_*(\xi) \rho_*(\zeta)\,a , \\
					 & = -\rho_*\left(  [\xi, \zeta]_{\text{\tiny{Lie$H$}}}+ \delta_{\xi\,} \zeta -\delta_{\zeta\,} \xi]\right)\,a, \\       
					 & = -\rho_*\big(\big\{ \xi, \zeta \big\}\big)\, a, \\
					 & = \delta_{\{ \xi,\, \zeta \}\,} a. 
\end{aligned}
\end{equation}	 
The local version of \eqref{Com-alpha}. 
Similarly, one finds: 
\begin{align}
[\delta_\xi, \delta_\zeta]\, A =\delta_\xi \delta_\zeta A - \delta_\zeta \delta_\xi A = D^A\big( \{ \xi, \zeta \} \big) = \delta_{\{ \xi,\, \zeta \}\,} A, 
 \end{align}
as the local counterpart of \eqref{Com-omega}
\medskip

The bracket \eqref{Gen-gauge-alg} is found in the physics literature, notably in the covariant phase space literature, as pointed out already.
It is often derived heuristically: Terms like $\delta_{\xi\,} \zeta$ are assumed to arise because of a field-dependence of the gauge parameters, i.e. $\zeta=\zeta(b)=\zeta(A, \upphi)$ -- itself arising either because of gauge fixing or because boundary conditions must be preserved. 
As we have seen, a field-dependence is not a necessary condition for an extended bracket to appear. 

Yet, as observed already, the geometric arena where field-dependent parameters $\zeta=\zeta(b)$ arise naturally is the field space $\Phi=\{b\}$ of a gauge theory seen as a principal bundle with structure group $\H_{\text{\tiny{loc}}}$. 
The~gauge group $\bs\Aut_v(\Phi)$ of $\Phi$, or its group of vertical automorphisms $\bs\Diff_v(\Phi)$,  are the geometric underpinning of the notion of field-dependent gauge transformations. Then, objects like $\zeta=\zeta(b)$ belong  to $C^\infty\big(\Phi, \text{Lie}\H_{\text{\tiny{loc}}} \big)\simeq\bs\diff_v(\Phi)$, and \eqref{Gen-gauge-alg}-\eqref{ext-bracket-local} can naturally be understood as arising from the Nijenhuis-Lie derivative and  Frölicher-Nijenhuis bracket on $\Phi$. See  \cite{Francois2023-a} for a detailed treatment of the case where $\Diff(M)$ is the structure group of $\Phi$.

\section{Conclusion} 

In this paper, we have detailed the global and local geometry arising from the group of vertical diffeomorphisms $\Diff_v(P)\simeq C^\infty(P, H)$ of a principal bundle $P$.  Notably, we have shown that its Lie algebra is realised  by the Nijenhuis-Lie derivative: This shows that the extended bracket often encountered in the gauge field theory literature -- mainly on gravity \cite{Bergmann-Komar1972, Salisbury-Sundermeyer1983} and its asymptotic symmetries (BMS and extensions)  \cite{Barnich-Troessaert2009, Gomes-et-al2018, Freidel-et-al2021, Freidel-et-al2021bis, Chandrasekaran-et-al2022, Speranza2022}  
 --  not only can be seen as the bracket on the sections of the action Lie algebroid associated to $P$,  
but also as an instance of the Frölicher-Nijenhuis bracket of vector-valued forms on $P$.

The action of $\Diff_v(P)\simeq C^\infty(P, H)$ on forms of $P$ defines generalised gauge transformations. They are distinct from standard gauge transformations induced by $\Aut_v(P)\simeq \H$, as can be seen when both are iterated. 
This implies  in particular that, on the base $M$, and contrary to local active gauge transformations (induced by $\H_{\text{\tiny{loc}}}$), local generalised active gauge transformations (induced by $C^\infty(U, H)$\footnote{Maps with unspecified transformations under either other elements of  $C^\infty(U, H)$ or of $\H_{\text{\tiny{loc}}}$.}) are a priori distinct from local gluings. 
Since the usual BRST framework \cite{Bertlmann}  indistinguishably encodes both infinitesimal gluings and  
Lie$\H_{\text{\tiny{loc}}}$, 
one may inquire as to the possibility that it must be adjusted to accommodate generalised gauge transformations.  
It is known that BRST cohomology is just the Chevaley-Eilenberg (CE) cohomology of Lie$\H \simeq \aut_v(\P)$ (or Lie$\H_{\text{\tiny{loc}}}$) with coefficients in  $\Omega^\bullet(P)$ (or local polynomials in the fields $\{b\}=\{A, \upphi, \cdots \}$) \cite{Bonora-Cotta-Ramusino, DeAzc-Izq}.
An extended BRST framework would probably just be the Lie algebroid cohomology of  $C^\infty(P, \text{Lie}H) \simeq \diff_v(P)$ with  coefficients in  $\Omega^\bullet(P)$. We may address this question elsewhere.

As we observed, most basic treatments of bundle geometry do not mention $\Diff_v(P)$, or elaborate much on it. 
A~good reason for this could be that, since its elements generically do no commute with the right action of the structure group $H$, these are generically not bundle automorphisms: i.e. they are not morphisms in the category of principal bundles. They~are  not ``relevant structures" from a categorical perspective. 
Indeed, the  action of $\Diff_v(P)$ on objects well defined for that category, such as the spaces of connections, equivariant and tensorial forms, is  ``problematic": it doesn't preserve those spaces. 

Still, $\Diff_v(P)$ maps a bundle to itself by preserving the fibration -- hence the well-defined action Lie groupoid and Lie algebroid viewpoint -- so the local picture is less troublesome. 
Indeed there is no issue for local gauge field theory, which can accommodate this extended notion of gauge transformations: A gauge argument still carries through, and the covariant derivative still does its job of preserving covariance under the action of $C^\infty(U, H) \supset \H_{\text{\tiny{loc}}}$. 
In a companion paper \cite{Francois2023-a}, we show how the the finite dimensional geometry exposed here may apply to the infinite dimensional bundle geometry of the field space $\Phi=\{b\}$ of a gauge theory: further clarifications of the literature mentioned above ensue.

\section*{Acknowledgment}  

This work was funded by the OP J.A.C MSCA grant, number CZ.02.01.01/00/22\_010/0003229, co-funded by the Czech government Ministry of Education, Youth \& Sports and the EU.
This research was also funded in part by the Austrian Science Fund (FWF), [P 36542].
Support from the service \emph{Physics of the Universe, Fields and Gravitation} at UMONS (BE) is also  acknowledged.
Finally, the author acknowledges the fruitful and clarifying feedback of an anonymous referee, which greatly improved the note.

{
\normalsize 
 \bibliography{Biblio8.5}
}

\end{document}